\documentclass{aa}  

\usepackage{graphicx}
\usepackage{txfonts}
\usepackage{color}
\usepackage{bm}
\usepackage{amsmath}

\usepackage{hyperref}
\hypersetup{
    colorlinks,
    linkcolor={blue},
    citecolor={blue},
    urlcolor={blue}
}
\newcommand{\newtext}[1]{\ifmmode\bm{ #1}\else\textbf{#1}\fi}

\begin{document}

   \title{Revisiting the picture of circumbinary disc truncation}

   \author{Enrico Ragusa\inst{\ref{unimi},\ref{unimiMat},\ref{ENS}},
   Elliot Lynch\inst{\ref{ENS},\ref{tokyo}},
   Guillaume Laibe\inst{\ref{ENS}},
   Richard Alexander \inst{\ref{leic}}}
   \authorrunning{E. Ragusa et al.}

   \institute{
       Dipartimento di Fisica, Università degli Studi di Milano, Via Celoria 16, 20133 Milano MI, Italy\label{unimi}\and
   Dipartimento di Matematica, Università degli Studi di Milano, Via Saldini 50, 20133, Milano, Italy \label{unimiMat}
       \and
   ENS de Lyon, CRAL UMR5574, Universite Claude Bernard Lyon 1, CNRS, Lyon, F-69007, France\label{ENS}
        \and        
    Department of Earth and Planetary Sciences, Institute of Science Tokyo, 2-12-1 Ookayama, Meguro, Tokyo 152-8551, Japan \label{tokyo}
    \and
    School of Physics and Astronomy, University of Leicester, Leicester, LE1 7RH, United Kingdom\label{leic}
    \\
   \email{enrico.ragusa@unimi.it}
             }

   \date{Received 6 February 2026; accepted 23 June 2026}
 
  \abstract
 {Discs surrounding binaries, referred to as circumbinary discs, are widely observed to develop central cavities carved by the gravitational influence of the binary.}
{Analytical estimates of cavity sizes predict truncation at $\sim 2 \textrm{--} 3$ binary separations, depending on binary properties. However, numerical studies show only qualitative agreement with these predictions: cavity sizes often evolve on long timescales and can exceed substantially the analytically predicted values. In this work, we revise this paradigm, suggesting that tidal truncation in circumbinary discs responds to additional dynamical parameters that have so far been neglected.}
{We analyse a suite of 80 numerical simulations of circumbinary discs to re-examine the physical mechanism responsible for cavity truncation and to provide a prescription for the cavity size independent of the state of evolution of the system.}
{We find that truncation depends not only on the binary parameters $a_{\rm bin}$, $e_{\rm bin}$, and mass ratio $q$, but also on the instantaneous cavity eccentricity $e_{\rm cav}$ and the relative apsidal orientation $\varpi_{\rm bin}-\varpi_{\rm cav}$. These quantities jointly determine the pericentre of the innermost stable disc orbit $R_{\rm p}$, in a way that shares some similarities with orbital stability in the restricted three body problem. Hydrodynamical effects introduce secondary corrections, with the disc scale height $H$ and viscosity $\alpha$ mildly shifting the cavity edge relative to the purely gravitational prediction. We introduce a semi-analytical prescription that captures these dependences for $R_{\rm p}$ and cavity semi-major axis $a_{\rm cav}$. Additional variability arises from resonant oscillations of the cavity, characterised by anti-phase evolution of $e_{\rm cav}$ and $a_{\rm cav}$, which are not captured by the model.}
{We conclude that cavity truncation for binaries with mass ratios $q>0.05$ is a process where the instantaneous orbital properties of the disc ($e_{\rm cav}$, $\varpi_{\rm cav}$) play a fundamental role and should be taken into account to accurately evaluate the truncation efficiency.}


   \maketitle
%
%-------------------------------------------------------------------

\section{Introduction}

In astrophysical environments, the material surrounding binaries -- here defined in the most general sense as two gravitationally bound masses -- is expected to organise into three disc structures: two small discs surrounding the individual masses of the binary, often referred to as circum-individual/circum-stellar discs (or sometimes mini-discs), and a large-scale disc surrounding the binary, referred to as a circumbinary disc.

The advent of the Atacama Large Millimeter Array (ALMA) has enabled spatially resolved imaging of circumbinary discs around several stellar binaries, revealing a wide range of morphological features -- including warps, spirals, azimuthal overdensities, and large material-depleted cavities -- that require detailed theoretical interpretation. Central cavities, in particular, are now exceptionally well-characterized in several benchmark systems. For instance, HD142527 \citep{biller2012,lacour2016,claudi2019} hosts a low-mass stellar companion, with well-constrained astrometry, and a bright asymmetric disc featuring a $\sim 90$ au cavity whose size appears inconsistent with the properties of the host binary \citep{nowak2024}. Similar cavity structures are observed in systems like V892 Tau \citep{long2021,alaguero2024}, IRAS 04158+2805 \citep{ragusa2021}, and GG Tau A \citep{cazzoletti2017,aly2018,keppler2020,toci2025}, all of which possess well-constrained orbital parameters that allow us to probe cavity-binary coupling. Beyond these primary examples, detections of resolved binaries embedded in circumbinary discs include L1448 IRS 3B \citep{tobin2016}, L1551 IRS 5 \citep{cruzsaenzdemiera2019}, BHB 2007-11 \citep{alves2019}, and IRAS 16293-2422 A \citep{maureira2020}.
While these systems primarily highlight radial depletion, others show evidence of strong three-dimensional effects: HD98800 exhibits a polar configuration between the disc and the binary orbit \citep{kennedy2019}, while GW Ori, another hierarchical triple, displays disc warping and tearing \citep{kraus2020}. 

Although these diverse architectures are primarily observed in young stellar systems, the underlying physics of binary-disc interactions is applicable across a much broader mass spectrum. Examples of discs interacting with gravitationally bound companions  range from Saturn's moons carving gaps in its rings (e.g. \citealp{goldreich1980}), planets carving gaps in protoplanetary disc observations (e.g. \citealp{long2018,andrews2018}, see \citealp{bae2023} for a review), to discs surrounding supermassive black hole binaries (e.g. \citealp{farris2014,ragusa2016,miranda2017,tiede2025}), which are crucial for determining the electromagnetic counterparts to black hole binary mergers.

The gravitational potential of the binary perturbs the surrounding disc and can drive a variety of dynamical responses, including spiral density waves (e.g., \citealp{rafikov2002a,fairbairn2022}), disc eccentricity $e_{\rm d}$ evolution (\citealp{teyssandier2016}), disc inclination evolution $i_{\rm d}$ (\citealp{aly2015,zanazzi2018}), and the formation of gaps and cavities through depletion of the disc surface density $\Sigma$ (\citealp{artymowicz1994,miranda2015}). This depletion process is commonly referred to as disc tidal truncation, that we will investigate in this paper. 

The transition between annular gaps and central cavities is set by the loss of stable orbits around the Lagrange points $L_4$ and $L_5$ for mass ratios $q \gtrsim 0.04$ \citep{murray1999}. Throughout this paper we focus on the cavity regime, therefore we restrict our analysis to binaries with $q>0.04$.
In this regime, the binary carves a material-depleted cavity whose size depends on both the properties of the binary and those of the disc. Linking the cavity size to system parameters therefore makes tidal truncation a powerful dynamical probe to indirectly constrain the properties of binaries and discs.

As discussed in more detail in Sec. \ref{sec:tidaltrunc}, the typical theoretical approach for estimating the cavity size is based on balancing tidal and viscous torques \citep{artymowicz1994,miranda2015}. By construction, this approach implies a progressive viscous evolution of the cavity towards a steady value over long timescales. Numerical simulations generally confirm this gradual evolution and the convergence to a steady cavity size (e.g., \citealp{hirsh2020,ragusa2020}), which, however, can be significantly larger than theoretical predictions \citep{thun2017,ragusa2020,penzlin2024}. In this paper, we propose a novel interpretation of tidal truncation, showing that besides the standard dependence on the binary orbital properties it also depends on the underlying kinematics of the disc, in particular on disc eccentricity and relative alignment of the disc and binary pericentres. These dependences share some similarities with those related to orbital stability in the restricted three-body problem, where the pericentre radius of the innermost stable orbit is related to the eccentricity of the test particle and orientation of the pericentre with respect to the binary. In particular, considering these additional dependences it is possible to retrieve the instantaneous size of the cavity with great accuracy, solving both the issue of gradual evolution of the cavity and the issue of its size exceeding the theoretical predictions.

This paper is the first in a series devoted to the evolution of circumbinary discs, with a particular emphasis on understanding how disc eccentricity shapes their dynamics. The remainder of this section (Sec. \ref{sec:tidaltrunc} and Sec. \ref{sec:eccdisc}) provides an overview of tidal truncation and disc eccentricity evolution in circumbinary discs. In Sec. \ref{sec:numsim} we introduce the numerical simulations used in this paper and throughout the series, describing the numerical setup and analysis methods. In Sec. \ref{sec:results} we present the main results, showing that tidal truncation is dynamically linked to disc eccentricity in a way that enables us to predict the instantaneous cavity size, and we provide a semi-empirical formula based on disc and binary properties. Finally, we discuss the implications of our findings in Sec. \ref{sec:discussion} and summarise our conclusions in Sec. \ref{sec:conclusion}.

\section{Circumbinary disc dynamics}

\subsection{Disc tidal truncation}\label{sec:tidaltrunc}

Binaries interact gravitationally with their circumbinary discs, exchanging angular momentum and energy that are transported through the disc in the form of waves. As mentioned in the previous section, this produces a variety of structures, such as spirals, azimuthal overdensities, gaps and cavities. These structures, in turn, exert a back reaction torque on the binary, altering its orbital properties.

The higher multipoles of the binary gravitational potential perturb the gas orbits through two concurring mechanisms: resonant and non-resonant.
Resonant truncation models interpret cavity formation as the result of angular momentum exchange at Lindblad resonances \citep{goldreich1980,artymowicz1994,miranda2015}. In this framework, the binary gravitational potential is decomposed into rotating perturbations that excite density waves at specific resonant radii. These waves transport angular momentum and energy through the disc, which are eventually deposited through viscous dissipation and shocks \citep{goodman2001,crida2006,cimerman2024}. By balancing resonant torques against viscous stresses, one can identify truncation radii associated with specific resonances (setting the cavity semi-major axis $a_{\rm cav}$), leading to truncation prescriptions that appear discretised, reflecting the underlying resonant structure. In the same resonant framework, cavity truncation  can be due to orbital instability that arises from resonance overlap, a phenomenon extensively studied in the context of the restricted three-body problem \citep{holman1999,quarles2018,adelbert2023,georgakarakos2024}.

Non-resonant truncation mechanisms instead attribute cavity formation to the destabilisation of disc orbits by the binary potential. Early work by \citet{papaloizou1977} and \citet{paczynski1977} already highlighted the role of tidal perturbations and orbital instability, with subsequent numerical studies confirming this picture \citep{pichardo2005,pichardo2008}. Two distinct non-resonant processes can be identified: (i) viscosity-dependent tidal torques arising from the global tidal wake, which lead to viscosity-independent truncation criteria (\citealp{papaloizou1977,artymowicz1994}\footnote{\citet{artymowicz1994} used it to calculate the truncation radius of circumbinary discs surrounding circular binaries, for which resonant truncation appears to underestimate the cavity size)}); and (ii) orbital intersection, where stable orbits intersect neighbouring trajectories, leading to shocks and material depletion in a fluid disc \citep{paczynski1977,rudak1981,pichardo2005,pichardo2008}.

Within these frameworks processes where orbital stability is the the driving mechanism of tidal truncation are expected to act on short, dynamical timescales, and for this reason we will refer to them as ``short-term processes''. Processes that rely on the balance between viscous and tidal torque are expected to require longer timescale to produce changes in the disc structure, most likely of the order of the local viscous timescale; for this reason we will refer to them as ``long-term processes''.

Despite methodological differences, existing truncation prescriptions display broadly consistent qualitative trends. In particular, the truncation radius increases with increasing binary or planetary eccentricity and mass ratio, until $a_{\rm cav}$ becomes only weakly dependent on $q$ for values of $q\gtrsim 0.2\textrm{--} 0.5$. Conversely, truncation becomes less efficient for increasing mutual inclination between the disc and the companion orbit, increasing disc aspect ratio $H/R$, and increasing disc viscosity $\nu$. These dependencies have been explored extensively in numerical studies of truncation in circumstellar and circumbinary discs for stellar companions and massive planets \citep{pichardo2005,pichardo2008,duffell2013,hirsh2020,ragusa2020,penzlin2024,dittman2024}, as well as for lower-mass planetary companions \citep{bryden1999,crida2006,duffell2013,rosotti2016,thun2017,facchini2018,zhang2018,thun2018,chen2021}.

\subsection{Circumbinary discs are eccentric}\label{sec:eccdisc}

Binary-disc interaction leads to disc eccentricity growth once a threshold mass ratio of $q\gtrsim 0.001$ is exceeded \citep{kley2006}, although numerical studies suggesting that larger values of $q$ may be required in 3D simulations \citep{li2023}. The eccentricity then increases until it reaches a saturation level, where eccentricity excitation and damping mechanisms balance. This saturation value depends not only on the binary and disc parameters, but also sensitively on the treatment of disc thermodynamics (e.g. \citealp{sudarshan2022, penzlin2025}). 

Eccentricity pumping due to the presence of companions has been extensively confirmed by numerical simulations, spanning a wide range of mass ratios. For low-mass companions, eccentricity growth has been measured in planet–disc interaction studies \citep{papaloizou2001,kley2006,dangelo2006,dunhill2013,ragusa2018,teyssandier2019,dempsey2021,tanaka2022,padgett2026}. At higher mass ratios, eccentric discs have been found both in circumbinary configurations \citep{macfadyen2008,marzari2009,shi2012,dunhill2015,miranda2017,ragusa2020,munoz2020,pierens2020,dittman2022,siwek2023,duffell2024} and in discs surrounding the individual components of binaries \citep{lubow1991a,lubow1991b,whitehurst1994,murray1996,kley2008,regaly2011}.
A large body of numerical work on circumbinary discs has consistently reported the formation of eccentric cavities \citep{dorazio2013,farris2014,ragusa2016,ragusa2017,price2018,calcino2019,calcino2020,heath2020,tiede2020,franchini2023,toci2025} or eccentric gaps in the presence of planetary or low-mass companions \citep{ataiee2013,zhu14b,scardoni2023,scardoni2025}. However, it is important to note that a more accurate treatment of disc thermodynamics -- including heating and cooling terms as well as in-plane cooling -- can, in some regimes, suppress or limit the growth of disc eccentricity \citep{penzlin2025}.

In the limit of a pressureless disc,  eccentric discs can be viewed as a set of nested, confocal elliptical orbits.
The morphology of an eccentric disc is fully described by its eccentricity profile\footnote{Often conveniently represented by the complex eccentricity vector $\mathcal E(a)=e_{\rm d}(a)\exp[i\varpi_{\rm d} (a)]$, which compactly encodes both the eccentricity amplitude and the orientation of the apsidal line.} $e_{\rm d}(a)$ and its pericentre longitude profile $\varpi_{\rm d}(a)$ (see \citealp{ragusa2024}). Hydrodynamical effects allow eccentricity to propagate as a wave, transporting through the disc the associated conserved quantity, the angular momentum deficit (AMD).
The wavelike nature of disc eccentricity leads to the existence of characteristic eccentricity profiles, known as eccentric eigenmodes, analogous to the energy eigenstates of a quantum mechanical potential well \citep{ogilvie2008,teyssandier2016}. Each $i$-th eccentric eigenmode has a characteristic profile $e^{\rm m}_{{\rm d},i}(a)$ and precesses rigidly, that is $\varpi_{\rm d}(a,t)= \dot \varpi^{\rm m}_{i}\times t$ with a characteristic precession rate $\dot \varpi^{\rm m}_{i}$, constant through the disc, determined by a combination of: i) pressure effects, that can induce either prograde or retrograde precession of the apsis, ii) the binary potential (that produces a prograde precession), iii) and disc self-gravity \citep{teyssandier2016,munoz2020,grcic2025}.

When the profile is not an eigenmode\footnote{For example, after a stellar flyby, or at the start of any circumbinary disc numerical simulation not explicitly initialised in an eigenmode.}, the disc evolution can be decomposed onto a superposition of nonlinearly-interacting modes with different amplitudes and precession rates. When this situation occurs, both precession and eccentricity evolution show peculiar oscillations similar to secular evolution observed in the context of planet-planet interactions in celestial mechanics (e.g. \citealp{ragusa2018,scardoni2023,scardoni2025}). Similarly, within the same simplified framework, it is reasonable to expect mean motion resonances between the disc and the binary to play a role in the modulation of disc eccentricity by introducing additional characteristic frequencies (e.g. \citealp{murray1999}).

Dissipative processes, such as viscosity, and dynamical effects, such as orbit crossing in the presence of steep eccentricity gradients, are expected to damp higher-order eccentric eigenmodes. As a result, on long timescales the disc is driven toward the fundamental eccentric mode, which acts as a natural attractor for the disc geometry. In this configuration the disc precesses rigidly at a constant rate while preserving a fixed eccentricity profile.
The physical mechanisms governing the damping of disc eccentricity, however, remain uncertain. Even if dissipation is assumed to be primarily viscous in origin the effect of viscosity on eccentricity is non-trivial: bulk viscosity tends to damp eccentricity (\citealp{ogilvie2001,goodchild&ogilvie2006,lynch2022b}), whereas shear viscosity can instead excite it, potentially leading to unstable eccentricity growth through viscous overstability (\citealp{syer1992,kley1993,lyubarskij1994,latter2006}).

Independently of the nature of dissipative effects, discs appear to be naturally equipped with a mechanism that leads to a progressive damping of their eccentricity over time. From an observational perspective, in the absence of sustained pumping mechanisms, that is in systems that are not circumbinary or showing no evidence of recent perturbations such as flybys \citep{cuello2019} or infall \citep{kuffmeier2023}, discs appear to evolve toward more circular configurations. While Class 0 discs are often observed to be asymmetric and eccentric (\citealp{ohashi2023}, and \citealp{commercon2024,adnan2026} for a theoretical picture), more evolved Class II discs are predominantly circular \citep{andrews2018,long2018}, suggesting a mechanism actively damps disc eccentricity. 

Collectively, these results indicate that, circumbinary discs are expected to reach a balance between  pumping and damping effects that drives the disc towards a non-circular quasi-steady state.
Despite this, eccentricity is often treated as a passive disc property, rather than a key dynamical ingredient. In the following sections we show that disc tidal truncation is tightly coupled to the evolution of disc eccentricity, to the extent that it correlates to the instantaneous value of the cavity semi-major axis.

\section{Numerical simulations}\label{sec:numsim}

We consider a total of 80 SPH simulations performed with the code \textsc{phantom} \citep{price2018a} over several years. Although carried out with different purposes, all simulations share the common feature of modeling circumbinary discs with variety of system parameters. In Sec. \ref{sec:setup} we present the general properties of the simulations, in Sec. \ref{sec:parspace} we discuss the parameter space and differences among them, and in Sec. \ref{sec:analysis} we outline the analysis performed.

\subsection{Setup}\label{sec:setup}
 
The simulations consist of two gravitationally bound masses $M_1$ and $M_2$, modeled as sink particles \citep{bate1995} with mass ratio $q=M_2/M_1$, semi-major axis $a_{\rm bin}=1$, eccentricity $e_{\rm bin}$ and total mass $M_{\rm bin}=M_1+M_2=1$, surrounded by a circumbinary accretion disc with a pre-carved cavity with initial semi-major axis $a_{\rm cav}^{\rm in}$, and total mass  $M_{\rm d}/M_{\rm bin}=\{5\times 10^{-3},10^{-4}\}$. For all the simulations we set the sink radius to $R_{\rm sink}=0.05 a_{\rm bin}$, which is generally smaller than one third of the Hill's radius of the lowest mass ratio we considered, $q=0.05$. The sink particles feel the back reaction force exerted by the disc. Although the low disc masses adopted lead to negligible evolution of the sinks, accounting for disc back-reaction has been shown to be an important ingredient for correctly capturing the system dynamics \citep{franchini2023}. We neglect disc self-gravity.

The disc is modelled using a number of SPH particles ranging from $N_{\rm part}=10^6$ to $5\times10^6$ (as detailed in Sec. \ref{sec:parspace}). This particle number is in general more than sufficient to fully spatially resolve the circumbinary disc (i.e., with a smoothing-length–to–scale-height ratio $h/H < 0.3$), but not the circumstellar discs within the gas-sparse cavity area (e.g. \citealp{duffell2024,david-cleris2025}), thereby preventing the formation of well-resolved circumstellar discs. In general, we note that key properties of circumbinary disc evolution, such as the growth rate of eccentricity, saturated disc eccentricity, and disc precession timescale, show good convergence with higher resolution simulations already at particle numbers of $N_{\rm part}=10^6$ \citep{duffell2024}. 

We initialise the disc with both circular orbits and eccentric orbits in order to study the impact of eccentricity on tidal truncation for values of eccentricity that are not naturally excited by the binary. In 3D simulations without additional perturbations (such as external perturbers, flybys, or infall), the cavity eccentricity of circumbinary discs typically remains below $e_{\rm cav} < 0.3$. The procedure used to initialise eccentric discs is described in Appendix \ref{appendix:eccentricdiscs}. The eccentricity profiles used and parameter choices are discussed in Sec. \ref{sec:parspace}.

We model disc viscosity using SPH artificial viscosity with $\alpha_{\rm AV}$ tuned following the prescription in \citet{price2018a} (i.e., viscous term applied to both approaching and receding particles) to reproduce an equivalent \citet{shakura1973} $\alpha$ viscosity, and with $\beta=2$ to prevent particle interpenetration. Such an implementation results both in an effective shear viscosity term and a spurious bulk viscosity term, and, for this reason, tends to be conservative regarding the efficiency of eccentricity excitation. 

The disc is assumed to be locally isothermal with $P=c_{\rm s}^2(R)\rho$, where $\rho$ is the local density, $c_{\rm s}(R)=c_{\rm s0}R^{-s}$ is the sound speed with $s=1/2$. In a circular disc, this choice yields a constant $H/R$ profile across the disc. The normalisation $c_{\rm s0}$ is chosen to match the desired $H/R$. Details about the values used for $\alpha$ and $H/R$ are discussed in Sec. \ref{sec:parspace}. We note that we deliberately choose to maintain a radial temperature profile despite the disc becoming 
eccentric: using a semi-major axis profile would imply the temperature profile to change following the disc geometry.  

The initial surface density profile\footnote{$\Sigma(a)$ is the ``circular'' surface density [defined below in Eq. (\ref{eq:sigma})], i.e. the surface density of a circular disc with the same mass distribution. It corresponds to $\Sigma^\circ$ in \citet{ogilvielynch2019} and $\Sigma_0$ in \citet{ragusa2024}, and should not be confused with the local surface density $\Sigma_{\rm loc}(a,\phi)$, depends on azimuth [see Eq. (\ref{eq:sigmaapp}) ].} is $\Sigma(a)=\Sigma_0(a/a_{\rm bin})^{-p}$, where 
$a$ is the disc semi-major axis, $p=1$ and $\Sigma_0$ tuned to get the corresponding $M_{\rm d}$. The definition of $\Sigma(a)$ for eccentric discs is discussed in Eq. \ref{eq:sigma} in Sec. \ref{sec:analysis}.

The simulations are evolved for different durations, spanning a wide range of final times, $150\,t_{\rm bin} \lesssim t_{\rm fin} \lesssim 5000\,t_{\rm bin}$, where $t_{\rm bin}$ is the binary orbital period (see below for further details). These timescales are generally sufficient to allow the system to relax from the initial conditions to a dynamical equilibrium, here meaning long enough that the relevant dynamical processes have fully developed, and, in most cases, to probe longer-term secular evolution -- that is, the system properties appear approximately constant over a few orbital periods, even if the disc undergoes secular evolution. We anticipate that our results will show that the cavity size is an “instantaneous” property of the system, determined solely by its current dynamical state. For this reason, any simulation evolved for at least a few dynamical timescales is adequate for our purposes.

\subsection{Parameter space}\label{sec:parspace}

The simulations explore the parameter space of the binary orbital properties: mass ratio $q=\{0.05, 0.06, 0.075, 0.1, 0.2, 0.5, 0.7, 1.0\}$, binary eccentricity $e_{\rm bin}=\{0.0, 0.1, 0.2, 0.3, 0.4, 0.5, 0.6, 0.7, 0.8\}$; and disc hydrodynamical properties $H/R=\{0.03, 0.04, 0.05, 0.055, 0.08, 0.1\}$ and \citet{shakura1973} viscous $\alpha$ parameter $\alpha=\{0.005, 0.01, 0.015, 0.05, 0.1\}$.
Simulations also explore different initial eccentricities $e_{\rm d}=\{0,0.3,0.5,0.65\}$. Details about the initialisation of eccentric discs with SPH can be found in Appendix \ref{appendix:eccentricdiscs}.
These simulations can be broadly divided into 4 main groups. In Appendix \ref{appendix:simpar} we report Tables \ref{tab:group1}, \ref{tab:group2}, \ref{tab:group3}, \ref{tab:group4} summarising the parameters and IDs of the simulations from each group. 

Group 1 (Tab. \ref{tab:group1}) consists of 22 simulations of initially circular circumbinary discs previously presented in \citet{ragusa2020}, modelled with $N_{\rm part}=10^6$ SPH particles, around circular binaries (with 2 exceptions: simulations 4F$e_0$ and 4FF$e_0$ that use $e_{\rm bin}\neq 0$). This set explores the binary mass ratio $q=\{0.05, 0.06, 0.075, 0.1, 0.2, 0.5, 0.7, 1.0\}$ and disc parameter space $H/R=\{0.03,0.05,0.1\}$ (one simulation, 4E2$e_0$ uses $H/R=0.055$) and $\alpha=\{0.005,0.01,0.1\}$. The initial cavity is pre-carved to $a_{\rm cav}^{\rm in}=2a_{\rm bin}$ and set the outer disc to $a_{\rm out}=7\,a_{\rm bin}$. The disc mass is $M_{\rm d}=0.005 \, M_{\rm bin}$ and simulations are typically evolved for $t_{\rm fin}\gtrsim 2500 \,t_{\rm bin}$, where $t_{\rm bin}$ is the binary orbital timescale, with a few exceptions that are generally evolved for at least $t_{\rm fin}\gtrsim 750\, t_{\rm bin}$.

Group 2 (Tab. \ref{tab:group2}): consists of 19 simulations of circumbinary discs, modelled using $N_{\rm part}=3\times 10^6$ SPH particles, expanding the parameter space by varying $e_{\rm bin}=\{0.,0.1,0.2,0.3,0.4,0.5,0.6,0.7,0.8\}$ at fixed $q=0.5$, combined with $H/R=\{0.05,0.1\}$ and $\alpha=\{0.005,0.05\}$. The disc mass is $M_{\rm d}=0.005 \, M_{\rm bin}$. The cavity is pre-carved to $a_{\rm cav}^{\rm in}=2a_{\rm bin}$ and the outer disc is $a_{\rm out}=10\,a_{\rm bin}$. Evolution timescales of different simulations vary with different disc aspect ratios, in particular: discs with $H/R=0.1$ are evolved for $t_{\rm fin}\gtrsim 1500 \,t_{\rm bin}$, while simulations with $H/R=\{0.05\}$ were evolved for $150 \,t_{\rm bin}\lesssim t_{\rm fin}\lesssim 600 \,t_{\rm bin}$.

Group 3 (Tab. \ref{tab:group3}): includes 33 simulations initialised with eccentric circumbinary discs, to study the evolution of the cavity under controlled conditions of disc eccentricity (labelled with IDs like XX$e_{\rm d}$). These simulations use $N_{\rm part}=5\times 10^6$ SPH particles and start with eccentric cavities $e_{\rm cav}=\{0.,0.3,0.5,0.65\}$ and constant eccentricity profile across the disc.
Among these, 8 simulations were initially started with pre-carved cavities with $a_{\rm cav}^{\rm in}=2 a_{\rm bin}$ and let evolve for a few tens of binary orbits (ID ending with $e_{\rm d,0}$), to determine the most appropriate value of $a_{\rm cav}^{\rm in}$ to be set to avoid creating an overdense cavity edge right after the simulation started. Then, 25 simulations were re-started with a cavity edge set to interpolate the value reached after $20t_{\rm bin}$ [i.e., $a_{\rm bin}^{\rm in}=a_{\rm cav}(t=20\,t_{\rm bin})$] for the appropriate $e_{\rm d}$, as per the values reported in Tab. \ref{tab:group3} (ID ending with $e_{\rm d}$). We note that this initialisation choice is meant only to avoid the abrupt evolution of the cavity size at the beginning of the simulation: as discussed later, the truncation radius for $e_{\rm cav}\sim 0.6$ is $a_{\rm cav}\sim 6a_{\rm bin}$, letting the simulation start from $a_{\rm cav}^{\rm in}=2a_{\rm bin}$, used for circular discs, is far from dynamical equilibrium.
The outer disc was set to $a_{\rm out}=15 a_{\rm bin}$. Evolution times in this group range $200 \,t_{\rm bin}\lesssim t_{\rm fin}\lesssim 1000\, t_{\rm bin}$.
We note that the simulations in Group 3 exhibit further evolution away from their initial conditions, indicating that our initialization does not artificially impose the cavity size at an arbitrary value.

Group 4 (Tab. \ref{tab:group4}):  6 simulations (labelled with IDs like XX$e_{\rm d}^{\rm m}$) use $N_{\rm part}=2\times10^6$ particles, and are initialized with $e_{\rm cav}(a_{\rm cav})=0.3$ and a decreasing eccentricity profile reproducing an eccentric eigenmode (e.g., \citealp{teyssandier2016, ogilvielynch2019}), meant to reproduce the long term spontaneous evolution of the disc to the fundamental eccentric eigenmode. The procedure to obtain such profiles is detailed in Appendix \ref{appendix:ecceigenmode}. Simulations use $a_{\rm in}=3\,a_{\rm bin}$ and $a_{\rm out}=24\,a_{\rm bin}$. Evolution times in this group are $t\gtrsim 1500\, t_{\rm bin}$.

\begin{figure*}
   \centering   \includegraphics[width=\textwidth]{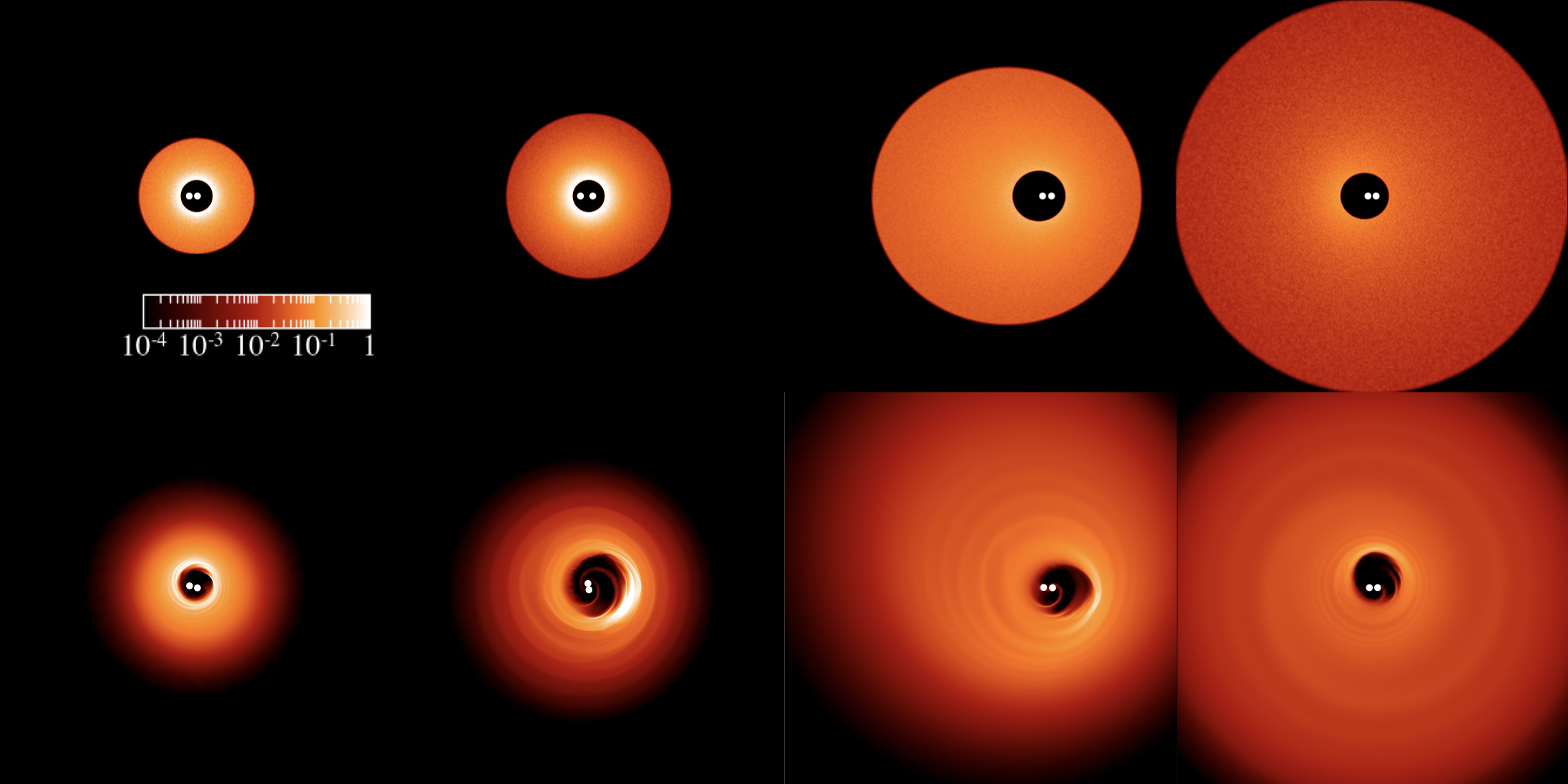}
   \caption{Examples of snapshots of local surface density distributions from numerical simulations with initially circular discs (representative of Group 1 and 2) and with initially eccentric discs (representative of Group 3 and 4). The top row shows the initial conditions. The bottom row an evolved disc after $t=300\, t_{\rm bin}$. 
   From left to right, Panel 1: shows simulation $2Ae_{\rm 0}$ ($q=0.1$, $e_{\rm bin}=0.$, $H/R=0.05$, $\alpha=0.005$, $e_{\rm d}=0$).
   Panel 2: shows simulation $6Be_{\rm b}$ ($q=0.5$, $e_{\rm bin}=0.5$, $H/R=0.05$, $\alpha=0.005$, $e_{\rm d}=0$).  Panel 3: simulation $1A3e_{\rm d}$ ($q=1$, $e_{\rm bin}=0.$, $H/R=0.08$, $\alpha=0.05$, $e_{\rm d}=0.3$). Panel 4: simulation $1B3e_{\rm d}^m$ ($q=0.5$, $e_{\rm bin}=0.5$, $H/R=0.04$, $\alpha=0.05$, $e_{\rm d}=0.3$). See Tabs. \ref{tab:group1}, \ref{tab:group2}, \ref{tab:group3}, \ref{tab:group4} for simulations IDs.}
\label{fig:simexample}%
\end{figure*}

We plot in Fig. \ref{fig:simexample} initial and evolved snapshots from representative simulations of Groups 1, 2, 3 and 4.

\subsection{Analysis}
\label{sec:analysis}

Each simulation of our dataset was analysed as follows. For each individual snapshot of the simulations, we extracted the eccentricity profile $e_{\rm d}(a)$ and longitude of pericentre profile $\varpi_{\rm d}(a)$, surface density profile $\Sigma(a)$, binary eccentricity $e_{\rm bin}$, the binary longitude of pericentre $\varpi_{\rm bin}$ and the binary semi-major axis. 

Disc profiles were obtained collecting SPH particles in $n_{\rm bin}=300$ semi-major axis bins spanning the range $R_{\rm in}^{\rm bins}=1.5a_{\rm bin}$ to $R_{\rm out}^{\rm bins}=15a_{\rm bin}$.
In eccentric discs, binning by radius, rather than by semi-major axis, can yield inaccurate results, as material at the same radius but at different azimuths belongs to different orbits.
The assignment of particles to bins was performed by computing the specific orbital energy of each particle,
\begin{equation}
    E_i=\frac{1}{2}\bm v_i^2 - \frac{GM_{\rm bin}}{R_i},
\end{equation}
where $\bm v_i^2$ is the dot product of the velocity vector of the i-th particle, and $R_{\rm i}$ is the distance from the centre of mass. Assuming gravitationally bound orbits, the corresponding semi-major axis was then obtained as
\begin{equation}
    a_i=-\frac{GM_{\rm bin}}{2E_i}.
\end{equation}
From this binning procedure we compute the circular surface density profile $\Sigma(a)$ as
\begin{equation}
\Sigma(a)=\frac{M_a(a)}{2{\rm \pi}a},\label{eq:sigma}
\end{equation}
where 
\begin{equation}
    M_a(a)=\nabla_a \sum_i^{a_i<a} m_i,
\end{equation}
that is the semi-major-axis gradient $\nabla_a$ of the cumulative disc mass enclosed within $a$, and $m_i$ is the mass of the individual particles.

To obtain $e_{\rm d}(a)$ and $\varpi_{\rm d} (a)$ we first calculate the complex eccentricity vector $\mathcal E_i=e_i\exp(i\varpi_i)$ of individual particles, where 
\begin{equation}
    e_i=\sqrt{1-\frac{h_i^2}{GM_{\rm bin}a_i}},\label{eq:ecc}
\end{equation}
where $h_i=x_iv_{y,i}-y_iv_{x,i}$ is the z-component of the specific angular momentum vector of the i-th particle, and 
\begin{equation}
    \varpi_i=\arctan2\left[-\frac{h_iv_{x,i}}{M_{\rm bin}}-\frac{y_i}{R_i},\frac{h_iv_{y,i}}{M_{\rm bin}}-\frac{x_i}{R_i}\right],
\end{equation}
as obtained from the expression for the complex eccentricity vector (e.g., \citealp{ogilvie2001}). The complex eccentricity vector is then averaged within each semi-major-axis bin to obtain $\mathcal E(a)$, from which we derive $e_{\rm d}(a)=|\mathcal E(a)|$ and $\varpi_{\rm d}(a)=\arg[\mathcal E(a)]$. 

This approach -- averaging the complex eccentricity vector -- removes the spurious eccentricity resulting from pressure effects \citep{teyssandier2017,commercon2024}. In particular, applying Eq. (\ref{eq:ecc}) to a particle on a circular orbit moving with sub-Keplerian velocity due to pressure support yields a spurious eccentricity $e_{i,{\rm spur}}\propto \sqrt{1/2}\,H/R$ with $\varpi=\theta_i+{\rm \pi}$, where $\theta_i$ the azimuthal coordinate of the particle, implying that the particle appears to lie at the apocentre of a fictitiously eccentric orbit. Averaging the complex eccentricity vector across all particles in a bin sums contributions with identical amplitude $e_{i,{\rm spur}}$ but phases uniformly distributed in azimuth, so that for a circular disc $\mathcal E(a)\approx 0$ even when the flow is sub-Keplerian. 

We recommend adopting an analogous semi-major-axis binning procedure when analysing grid-based simulations (as done for example by \citealp{teyssandier2017,commercon2024}).  This issue can be straightforwardly addressed by computing the semi-major axis of individual grid cells and binning them in the same manner as SPH particles.

We define the cavity semi-major axis (or cavity size) the location $a_{\rm cav}$ where \begin{equation}
    \Sigma(a_{\rm cav})=\frac{1}{2}\max_a[\Sigma(a)],\label{eq:acavdef}
\end{equation} 
that is the location in the disc where the disc surface density becomes $50\%$ of its maximum value at the cavity edge. The cavity eccentricity and cavity longitude of pericentre are then defined as $e_{\rm cav}=e_{\rm d}(a_{\rm cav})$ and $\varpi_{\rm cav}=\varpi_{\rm d}(a_{\rm cav})$, respectively. 
For completeness, we define the semi-major axis of the density maximum $a_{\rm cav}^{\rm max}$ the location where
\begin{equation}\Sigma(a_{\rm cav}^{\rm max}) = \max_a [\Sigma(a)]; \label{eq:acavmax}\end{equation} 
similarly, we define the eccentricity and pericentre longitude at the density maximum as $e_{\rm cav}^{\rm max} = e(a_{\rm cav}^{\rm max})$ and $\varpi_{\rm cav}^{\rm max} = \varpi(a_{\rm cav}^{\rm max})$, respectively. 

Finally, we compute the cavity precession rate by ``unwrapping'' the value of $\varpi_{\rm cav}(t)$ to obtain a monotonic (i.e. non-periodic) phase; we apply a linear interpolation of $\varpi_{\rm cav}(t)$ and smooth it with a time window $\Delta t=40 \,t_{\rm bin}$ to suppress short term variability; we finally take the time derivative to obtain $\dot \varpi_{\rm cav}(t)$.

\section{Results}\label{sec:results}

\begin{figure*}
   \centering  
   \includegraphics[width=\textwidth, trim=0.6cm 1.73cm 0.6cm 0.55cm, clip]{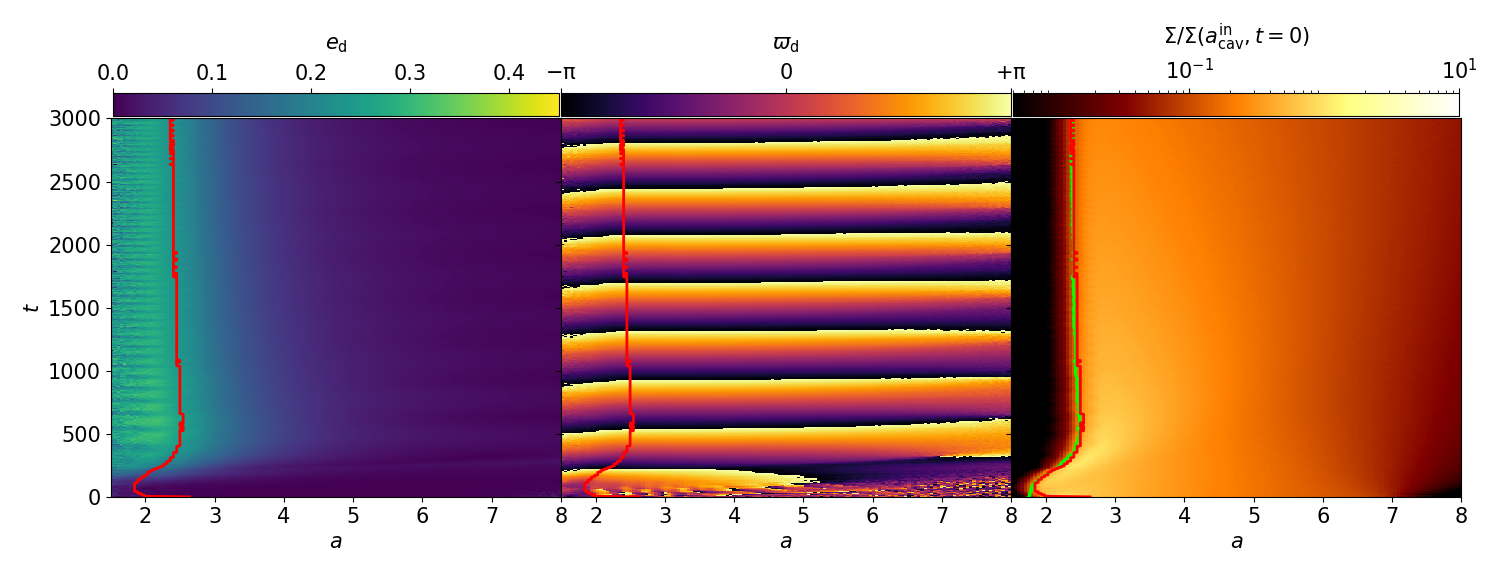}\\

   \includegraphics[width=\textwidth, trim=0.25cm 1.73cm 0.6cm 2.92cm, clip]{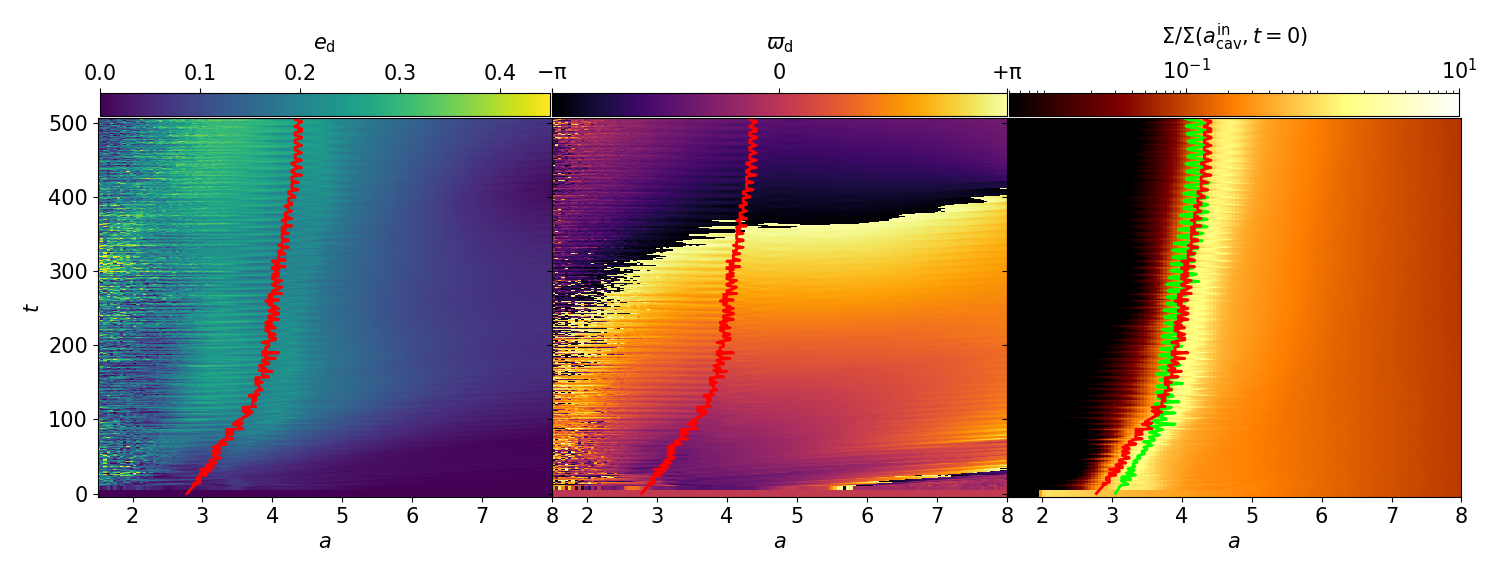}\\
  
   \includegraphics[width=\textwidth, trim=0.25cm 1.73cm 0.6cm 3cm, clip]{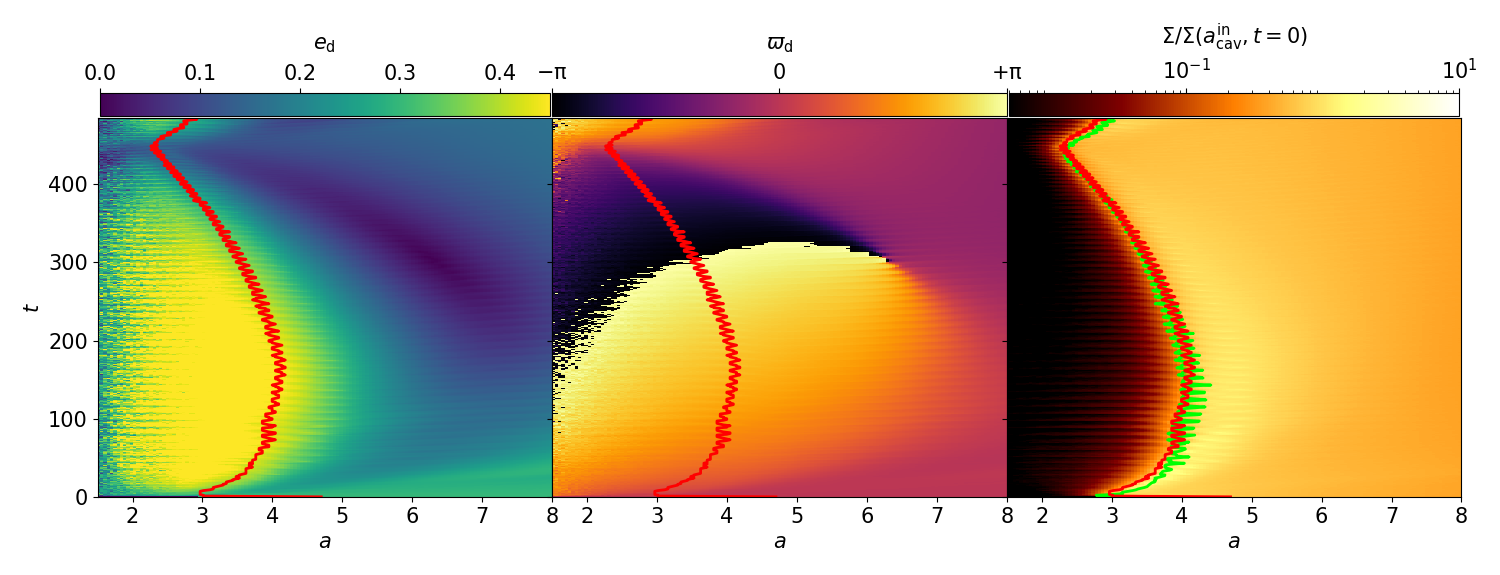}\\
   \includegraphics[width=\textwidth, trim=0.6cm 0.7cm 0.6cm 3cm, clip]{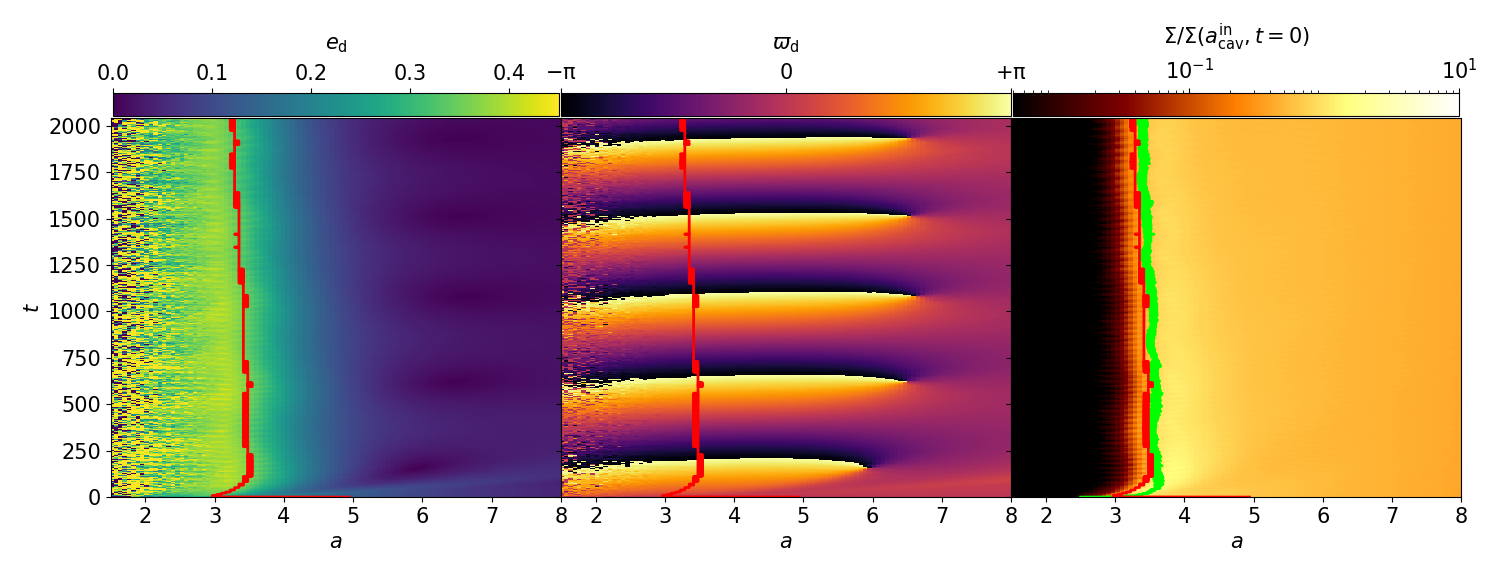}
   \caption{Evolution colourplots of $e_{\rm d}(a)$, $\varpi_{\rm d}(a)$, $\Sigma(a)$ for representative simulations in Fig. \ref{fig:simexample}. Simulation $2Ae_0$ from Group 1 (first row, $q=0.1$, $e_{\rm bin}=0.$, $H/R=0.05$, $\alpha=0.005$, $e_{\rm d}=0$), simulation $6Be_{\rm b}$ from Group 2 (second row, $q=0.5$, $e_{\rm bin}=0.5$, $H/R=0.05$, $\alpha=0.005$, $e_{\rm d}=0$), simulation $1A3e_{\rm d}$ from Group 3 (third row, $q=1$, $e_{\rm bin}=0.$, $H/R=0.08$, $\alpha=0.05$, $e_{\rm d}=0.3$) and $1B3e_{\rm d}^{\rm m}$ form Group 4
  (fourth row, $q=1$, $e_{\rm bin}=0.$, $H/R=0.04$, $\alpha=0.01$, $e_{\rm d}=0.3$). Each plot shows the typical temporal evolution of the disc profiles, where each horizontal ``slice'' of the plot shows the quantity profile at time $t$ in units of $t_{\rm bin}$:  eccentricity $e_{\rm d}(a)$ (left panel), pericentre longitude $\varpi_{\rm d}(a)$ (middle panel) and surface density $\Sigma_{\rm d}(a)$ (right panel) throughout the simulation.  Red lines mark the location of the cavity edge $a_{\rm cav}$ defined as the location were the surface density value is half of that of the maximum (Eq. \ref{eq:acavdef}). The green line in the surface density plot shows the theoretical prediction obtained using Eq. \ref{eq:acav}, as discussed in Sec. \ref{sec:avsecc}. As the  simulations start, the disc eccentricity grows abruptly and undergoes prograde precession: rigid for simulation $2Ae_0$, $6Be_{\rm b}$ and $1B3e_{\rm d}^{\rm m }$, and differential for simulation $1A3e_{\rm d}$. See Tabs. \ref{tab:group1}, \ref{tab:group2}, \ref{tab:group3} for simulations IDs.}
\label{fig:simevo}%
\end{figure*}

\subsection{Circumbinary disc evolution}

\begin{figure*}
   \centering   
   \includegraphics[width=\columnwidth]{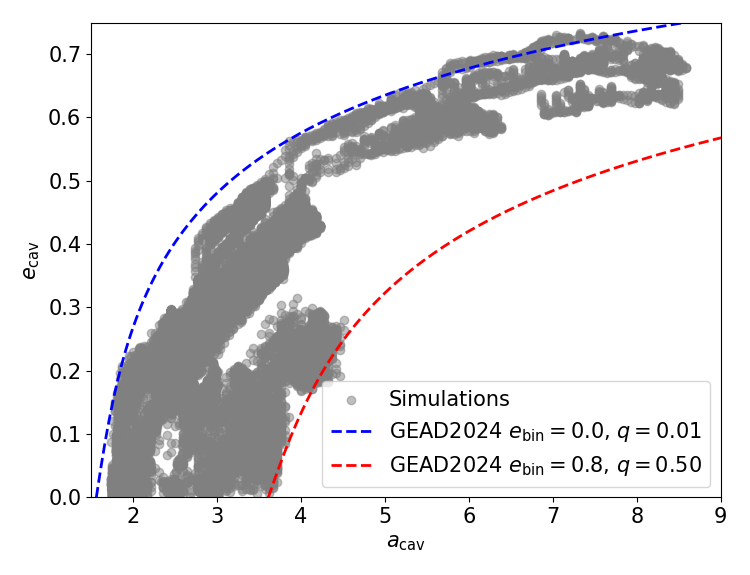}
   \includegraphics[width=\columnwidth]{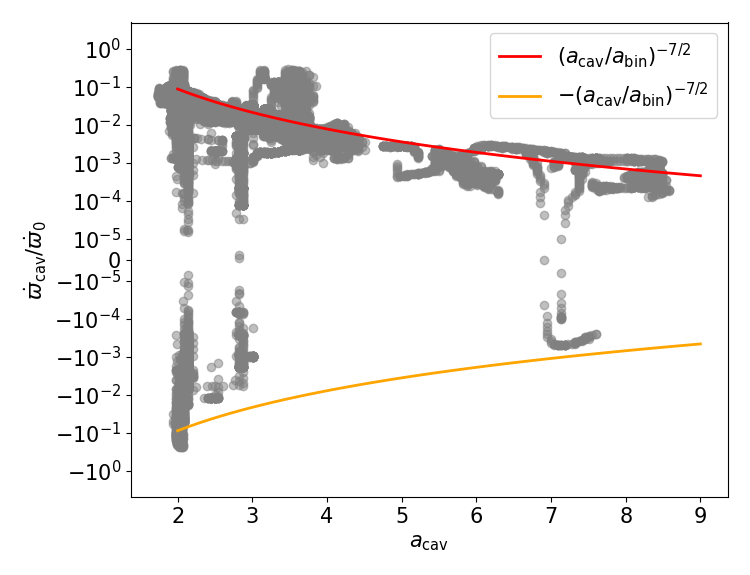}
 
   \caption{General dependence of cavity eccentricity $e_{\rm cav}$ and precession rate $\dot \varpi_{\rm cav}$ as a function of cavity semi-major axis for the full simulation dataset (see also Fig. \ref{fig:acavecavq1}, \ref{fig:dataset2} and \ref{fig:datasetNbody} for dependences on the simulation parameters); each data point represent a simulation snapshot. Left panel: Cavity eccentricity $e_{\rm cav}$ as a function of cavity semi-major axis $a_{\rm cav}$ measured from the simulations. The dashed lines show theoretical predictions for the innermost stable orbit surrounding binaries from \citet{georgakarakos2024}, shown as qualitative boundaries of the chaotic region in the 3-body problem for our parameter space using $q=0.01$ and $e_{\rm bin}=0$ (dashed, blue line) and $q=0.5$ and $e_{\rm bin}=0.8$ (dashed, red line), as reported in the legend. Right panel: Normalised precession rates $\dot \varpi_{\rm cav}/\varpi_0$ using $\dot \varpi_0$ in Eq. (\ref{eq:varpi0}); the red and orange solid lines show relations $\pm(a_{\rm cav}/a_{\rm bin})^{-7/2}$, respectively.}
\label{fig:dataset}%
\end{figure*}

In Fig. \ref{fig:simevo} we show examples of disc evolution from the selected representative simulations from Group 1, 2, 3 and 4 that we showed in Fig. \ref{fig:simexample}.
The simulations considered in this figure show the quantities $e_{\rm d}(a)$ (left panels), $\varpi_{\rm d}(a)$ (central panels), and $\Sigma(a)$ (right panels) as a function of time $t$ and disc semi-major axis $a$. These plots show profiles of the quantities mentioned above (colour) as a function of semi-major axis (x-axis), across the time span of the simulation (y-axis). In other words, each horizontal cut shows the profile of $e_{\rm d}(a)$, $\varpi_{\rm d}(a)$, $\Sigma(a)$ at time $t$. 

The typical evolution we observe in all simulations is an abrupt change of the disc eccentricity $e_{\rm d}$ until it reaches a saturation value (which varies between $e_{\rm cav}=0.05 \textrm{--}0.4$), accompanied by the growth of the cavity size $a_{\rm cav}$, and by the precession of the disc longitude of pericentre $\varpi_{\rm d}$, occurring on a timescale of the order of hundreds of orbits. 

Disc eccentricity grows exponentially, as soon as a ``seed'' of disc eccentricity is set by the dynamical perturbative effects of the binary on the disc dynamics. This behaviour is consistent with previous theoretical (e.g., \citealp{ogilvie2001,teyssandier2016}) and numerical studies (e.g., \citealp{miranda2017,munoz2018,ragusa2020,duffell2024}). 

Simulations from Group 1 with circular binaries and initially circular cavities remain circular for a few hundred orbits and then exhibit a rapid growth of both disc eccentricity and the cavity size; the final values of disc eccentricity $e_{\rm cav}$ depend on $q$: the higher $q$, the more eccentric the disc; for the highest mass ratios ($q=\{0.5;0.7;1\}$) the final eccentricity saturates to $e_{\rm cav}\approx 0.35$ independently of $q$ (see also \citealp{ragusa2020}). 

Simulations in Group 2 exhibit eccentricity growth as soon as the simulations start, with final values ranging from $e_{\rm d}\approx 0.1$ to $e_{\rm d}\approx 0.3$ depending on the disc aspect ratio ($H/R=\{0.05;0.1\}$): thinner discs become more eccentric (see also Sec. \ref{sec:discussion} for further discussion).

Simulations in Group 3, that are initialised with a constant eccentricity profile (IDs of type $XXe_{\rm d}$), show further growth of disc eccentricity as soon as the simulations start, while those in Group 4, initialised with an eccentric eigenmode (IDs of type $XXe_{\rm d}^{\rm m}$), show a quasi-steady, slowly decreasing, eccentricity at the edge of the cavity.

The evolution of $e_{\rm d}(a)$, $\varpi_{\rm d}(a)$, and cavity size $a_{\rm cav}$ (through $\Sigma(a)$ plots), described here, is shown for representative simulations from Group 1, 2, 3 and 4 in Fig. \ref{fig:simevo} and summarised for all other simulations in Fig. \ref{fig:dataset}, where we show $e_{\rm cav}$ and $\varpi_{\rm cav}$ as a function of the cavity size $a_{\rm cav}$. The general dependence of $a_{\rm cav}$ on the different simulation parameters will be further discussed in Sec. \ref{sec:dependences} and Sec. \ref{sec:discussion}.

After $\sim 200\,t_{\rm bin}$, we observe the onset of rigid precession of the disc up to $a\approx 8\,a_{\rm bin}$ in Group 1, Group 2, and Group 4. In contrast, simulations with constant eccentricity profiles in Group 3 show differential precession: that is, different semi-major axis annulii precess at different rates. This twists the disc and produces a characteristic $m=1$ spiral overdense feature, due to eccentric orbit clustering (e.g., see also \citealp{ragusa2024,romanova2024}). 
The central panels of Fig. \ref{fig:simevo} exemplify three cases of rigid precession ($2Ae_0$, $6Be_{\rm b}$ and $1B3e_{\rm d}^{\rm m }$) and one of differential precession in our dataset ($1A3e_{\rm d}$). 

As discussed in Sec. \ref{sec:eccdisc}, a longitude of pericentre profile precessing rigidly is typically associated with a dominant eccentric eigenmode that drives the main pericentre drift (e.g. \citealp{teyssandier2016}). Such a behaviour is typical for simulations that evolve for a sufficiently long timescale for the system to relax to the fundamental eigenmode (e.g. \citealp{miranda2017,thun2017,ragusa2018,munoz2020,penzlin2024,dittman2024}). In contrast, a differentially precessing disc is associated with the co-existence of multiple eigenmodes, and consequently it is most commonly observed in Group 3 simulations starting with a constant eccentricity profile (i.e., far from the fundamental eigenmode). 

All simulations exhibit prograde ($\dot \varpi>0$) precession of the longitude of pericentre. However, as shown in Fig. \ref{fig:dataset}, a few simulations have precession rates transitioning to retrograde ($\dot \varpi<0$). These cases are essentially from Group 3, but they do not necessarily result in a physical long term retrograde drift of the pericentre. In particular, we identify three cases: i) simulations 01B6$e_{\rm d}$ and 1B6$e_{\rm d}$ show a periodic modulation of the precession rate such that it slows down and makes a brief transition to retrograde superimposed to an otherwise prograde precession rate; ii) simulations 1B0$e_{\rm d}$, 05B0$e_{\rm d}$ present a negative precession rate in the early stages of the simulations, that we interpret as artifacts related to the difficulty of assigning a well-defined angle when $|\mathcal E|\sim 0$; iii) simulation 01A3$e_{\rm d}^{\rm m}$ transitions to constant retrograde precession after $t\sim 1700\, t_{\rm bin}$, and is the only simulation in the dataset characterised by real, long-term retrograde drift of the pericentre. 

While a full investigation of this transition is beyond the scope of this paper, it may be explained by the coexistence of multiple eccentric eigenmodes with positive and negative precession rates, as previously observed in \citet{ragusa2018}: initially, prograde eigenmodes dominate; retrograde precession occurs if a retrograde eigenmode becomes dominant, either due to damping of the prograde mode or resonant excitation of the retrograde mode.

More generally, the precession rates of the simulations appear to follow the dependence 
\begin{equation}
\dot \varpi_{\rm cav}\approx\dot \varpi_{\rm Q}=\dot \varpi_0\left(\frac{a_{\rm cav}}{a_{\rm bin}}\right)^{-7/2},\label{eq:varpi}
\end{equation}
where $\dot \varpi_0$ is:
\begin{equation}
  \dot \varpi_0=\frac{3}{4}\frac{ q_{\rm bin}}{(q_{\rm bin}+1)^{2}}\frac{1+\frac{3}{2} e_{\rm bin}^2}{(1-e_{\rm cav}^2)^2}\frac{2{\rm \pi}}{t_{\rm bin}}.\label{eq:varpi0}
\end{equation}
The rate $\varpi_{\rm Q}$ corresponds to the theoretical prediction for the precession rate from the quadrupole potential from \citet{thun2017}, based on the analytical framework of \citet{moriwaki2004}.

Overall, the datapoints in Fig. \ref{fig:dataset} appear to show the $a^{-7/2}$ dependence expected from Eq. (\ref{eq:varpi}). 
This result is consistent with the findings by \citet{munoz2020} and \citet{grcic2025} that the precession rate varies between $0.2\,\dot \varpi_{\rm Q}\lesssim \dot \varpi_{\rm cav}\lesssim \dot \varpi_{\rm Q}$ also when pressure effects dominate the precession rate over the quadrupole potential terms.

\subsection{The evolution of tidal truncation efficiency}\label{sec:avsecc}

Simulations presented in this work show instantaneous variations of the cavity size $a_{\rm cav}$ in response to the dynamical evolution of the system. In our simulations, the cavity size spans the range $2\,a_{\rm bin}\lesssim a_{\rm cav}\lesssim 8\,a_{\rm bin}$ (as visible in Fig. \ref{fig:dataset}), depending on the system parameters. 

In this section, we analyse the evolution of tidal truncation as a function of the orbital properties of circumbinary discs. 
Our results highlight the need of introducing dynamical variables that were not previously considered in existing theoretical models in order to capture the details of the evolution of the cavity size. We show that the introduction of these new variables improves the estimate on the cavity size to the extent that it is possible to match the value of the cavity size of a numerical simulation with a semi-analytical prediction. Such a semi-analytical model considers the instantaneous dynamical properties of the disc and binary, and ignores the disc stage of evolution or achievement of a steady state.

\begin{figure}
   \centering   \includegraphics[width=\columnwidth]{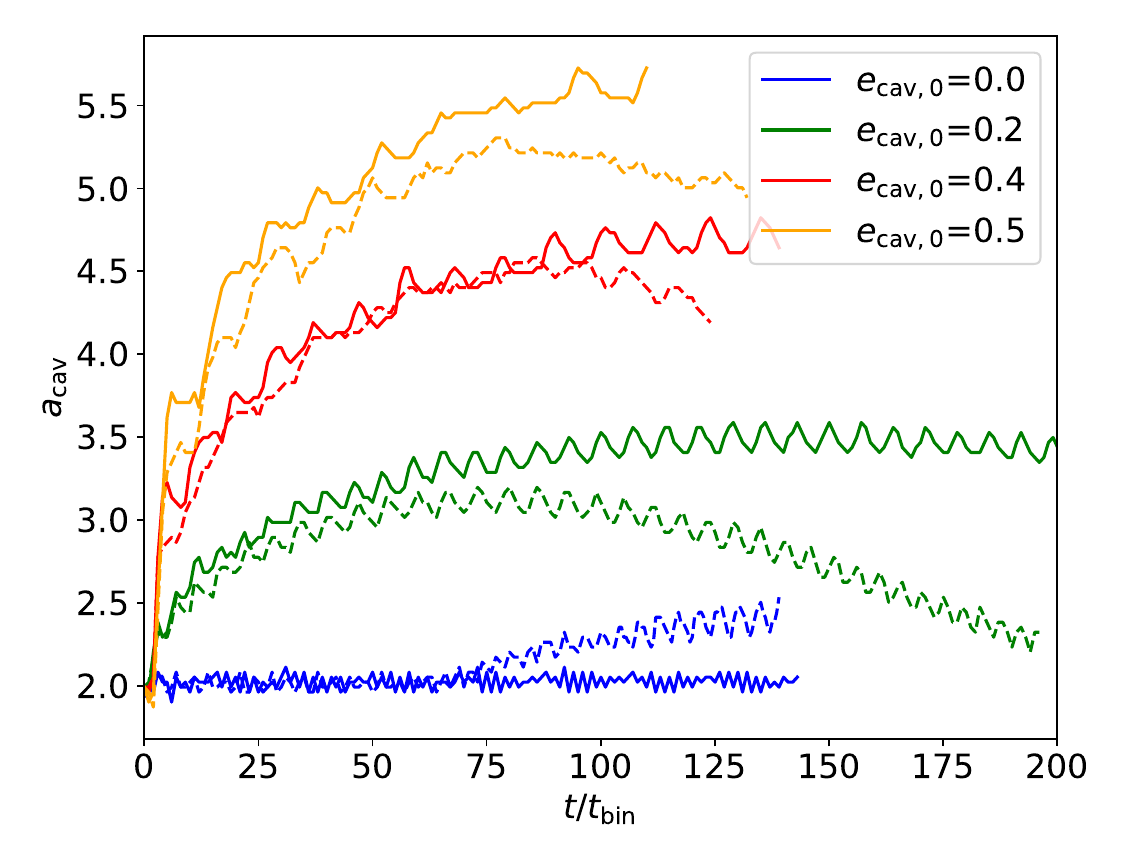}\\
   \includegraphics[width=\columnwidth]{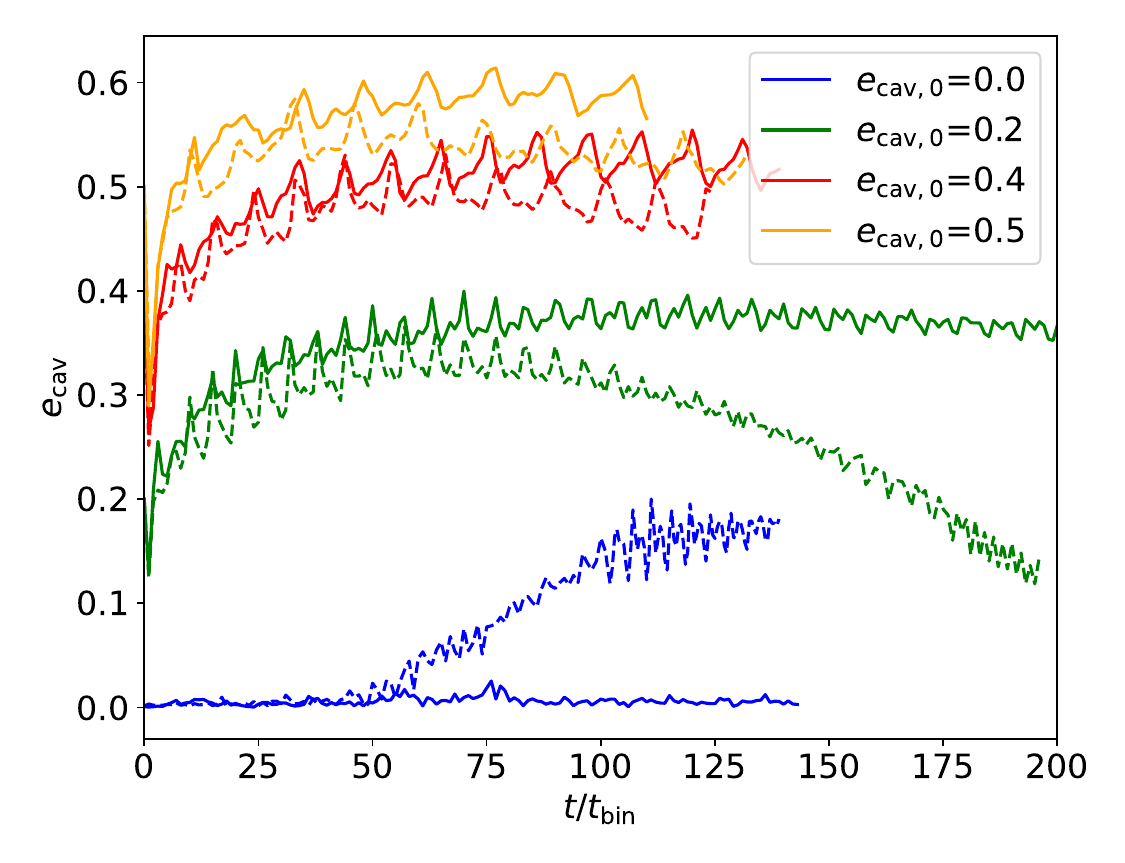}\caption{Initial evolution of $a_{\rm cav}$ (top panel) and $e_{\rm cav}$ (bottom panel) as a function of time for Group 3 simulations with initial disc eccentricity $0\leq e_{\rm cav,0}\leq 0.5$: that is, simulations 1A(0-5)$e_{\rm d,0}$ ($H/R=0.1$, solid lines) and 1B(0-5)$e_{\rm d,0}$ ($H/R=0.05$, dashed lines) and undersized cavity ($a_{\rm cav}^{\rm in}=2a_{\rm bin}$, IDs ending with $e_{\rm d,0}$ in Tab. \ref{tab:group3}). Different colours are different initial disc eccentricities as indicated in the legend. For reference, the local viscous timescale for the simulation is $t_\nu\sim 200 \,t_{\rm orb}\gtrsim 800\,t_{\rm bin}$ and $t_\nu\sim 800 \,t_{\rm orb}\gtrsim 3200 t_{\rm bin}$ for $H/R=0.1$ and $H/R=0.05$ respectively (for a cavity size $a_{\rm cav}\gtrsim2.5\,a_{\rm bin}$ the orbital timescale is $t_{\rm orb}\gtrsim 4 t_{\rm bin}$).}
\label{fig:dyntime}%
\end{figure}

\subsubsection{Short timescales and long timescales} 

In Fig. \ref{fig:dyntime}, we track the evolution of a few Group 3 simulations starting with an undersized initial cavity ($a_{\rm cav}^{\rm in}=2a_{\rm bin}$, ID ending with $e_{\rm d,0}$ in Tab. \ref{tab:group3}) for different initial values of disc eccentricity. For values of $e_{\rm d}> 0$ (for $e_{\rm bin}=0$ the cavity has already its equilibrium size) the cavity shows an abrupt expansion beyond its initial size. For example, for $e_{\rm cav,0}=0.5$ the cavity size grows from $a_{\rm cav}=2\,a_{\rm bin}$ to $a_{\rm cav}\approx 3.8\,a_{\rm bin}$ in less than $10$ binary orbits. This initial growth occurs on a timescale that is significantly shorter than the viscous timescale across the extent $\Delta R\sim H$ density gradient at the cavity edge $\tau_\nu\sim (\Delta R/R) [\alpha (H/R)^2]^{-1}\times t_{\rm orb}\sim 200 \textrm{--}800 \,t_{\rm orb}$, where $t_{\rm orb}$ is the local orbital timescale in the disc ($t_{\rm orb}\gtrsim 4t_{\rm bin}$ at the cavity edge). The simulations also show an evolution that is independent of viscosity ($\nu\propto\alpha (H/R)^2$), as it initially shows no apparent difference between $H/R=0.05$ and $H/R=0.1$ runs, despite a factor $4$ difference in $t_\nu$ between the two $H/R$ regimes. The cavity sizes show a strong correlation with the cavity eccentricity. After the initial transient, in most simulations shown in Fig. \ref{fig:dyntime} the cavity appears to progressively relax towards a quasi-steady configuration with short-period oscillations ($\Delta t\lesssim 10\,t_{\rm bin}$) of eccentricity and cavity size, superimposed to a steady trend reached across timescales $\Delta t\gtrsim 100-200\,  t_{\rm bin}$. Two exceptions are blue and green dashed lines ($e_{\rm d,0}=0$ and $e_{\rm d,0}=0.2$, respectively, both using $H/R=0.05$) that appear not to show the achievement of a steady configuration.

From now on, we will refer to those processes producing effects on the cavity evolution over timescales $\Delta t\lesssim10\,t_{\rm bin}$ as short-term processes and those that produce effects on timescales $\Delta t\gtrsim 100-200\,  t_{\rm bin}$ as long-term processes.

\subsubsection{Comparison with N-body results}\label{sec:compNbody}

As discussed in Sec. \ref{sec:tidaltrunc}, the gravitational potential of the central binary can be decomposed into a series of Fourier harmonics, each corresponding to a specific resonance location in the surrounding disc. As the test particle eccentricity increases (i.e., $e_{\rm cav}$), the widths of these individual resonance islands expand. When the perturbation is sufficiently strong, these islands overlap. This overlap creates a region where gas orbits become chaotic and are rapidly cleared, effectively setting the inner boundary of the circumbinary cavity.
In this framework, the pericentre of the innermost stable orbit, $R_{\rm p}^{\rm crit}$ serves as a representative boundary beyond which the phase space becomes dominated by unstable trajectories, effectively truncating the gaseous disc. In this framework, when compared to a fluid disc, $R_{\rm p}^{\rm crit}$ marks the location where the density profile begins to decrease, thereby setting the semi-major axis of the density maximum $a_{\rm cav}^{\rm max}$ (Eq. \ref{eq:acavmax}). It is important to note that a proper comparison between fluid simulations and N-body stability limits requires using this density peak ($a_{\rm cav}^{\rm max}$) rather than the half-maximum location ($a_{\rm cav}$) used elsewhere in this work. 

We note that the observed correlation between $a_{\rm cav}$ and $e_{\rm cav}$ shares some similarities with the stability criteria of particle orbits within the restricted three-body framework just described, offering an interesting celestial mechanics analogy for the gas behavior.
A first qualitative hint of this dynamical behaviour can be noticed in Fig. \ref{fig:dataset}, where the dashed lines mark the predicted extent of the chaotic orbit region in the restricted three-body problem for the extremes of our parameter space using the prescription of \citet{georgakarakos2024} (see their Eq. 10). These lines define the stability limit $a^{\rm crit} = R_{\rm p}^{\rm crit}/(1-e)$ for a test particle (\citealp{holman1999,quarles2018,adelbert2023}). Notably, already the $e_{\rm cav}$ and $a_{\rm cav}$ values (i.e. half-maximum values) from our simulation snapshots lie consistently within these theoretical curves, and qualitatively follow the characteristic trend $a_{\rm cav}\sim (1-e_{\rm cav})^{-1}$ of three-body orbital stability.

A more quantitative comparison is illustrated in Fig. \ref{fig:acavecavq1} and \ref{fig:datasetNbody} (shown in Appendix \ref{appendix:simpar}  due to space constraints, additional discussion can be found there) which compares the measured values of $a_{\rm cav}^{\rm max}$ and the predicted $R_{\rm p}^{\rm crit}$ by \citet{georgakarakos2024}. By specifying the values of $e_{\rm bin}$, $q$, and $e_{\rm cav}^{\rm max}$, one can reproduce the location of $a_{\rm cav}^{\rm max}$ across our dataset, proving that the N-body prescription constitutes a reasonable initial approximation for the disc truncation mechanism. 

Our dataset suggests that for a fixed value of $e_{\rm cav}$ a range of $a_{\rm cav}$ values are possible. However, we note that the stable region in the $a_{\rm cav}-e_{\rm cav}$ plane is characterised by a complex, ``ragged'', sawtooth profile. This makes the region where the transition between stable and unstable orbits occurs more similar to a ``band'' rather than a hard threshold. Furthermore, we cannot exclude the contribution to this process of long-term processes, such as the action of individual second order resonances. These resonances, strengthened by the larger disc eccentricity, could contribute on longer timescales to the process of tidal truncation via tidal-viscous torque balance, producing the observed dispersion of values of $a_{\rm cav}$.

We speculate that this agreement with N-body stability suggests that resonance overlap, and resulting chaotic orbits in the cavity region, actively participate to the tidal truncation process on short-timescales.
On longer timescales, the growth of the disc eccentricity make some eccentric individual resonances stronger, altering the viscous-tidal torque balance, and impacting on the long-term evolution of the system. However, the current data do not allow us to disentangle whether tidal truncation operates via an interplay of these short-term processes and long-term processes, or whether they act as mutually exclusive mechanisms.

\subsubsection{Hydrodynamic and viscous corrections}\label{sec:hydrvisc}

\begin{figure}
   \centering   \includegraphics[width=\columnwidth]{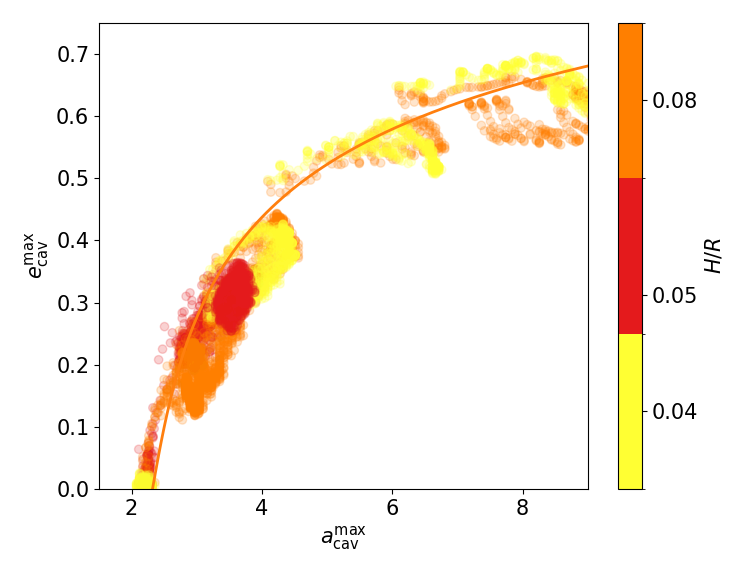}\\
   \includegraphics[width=\columnwidth]{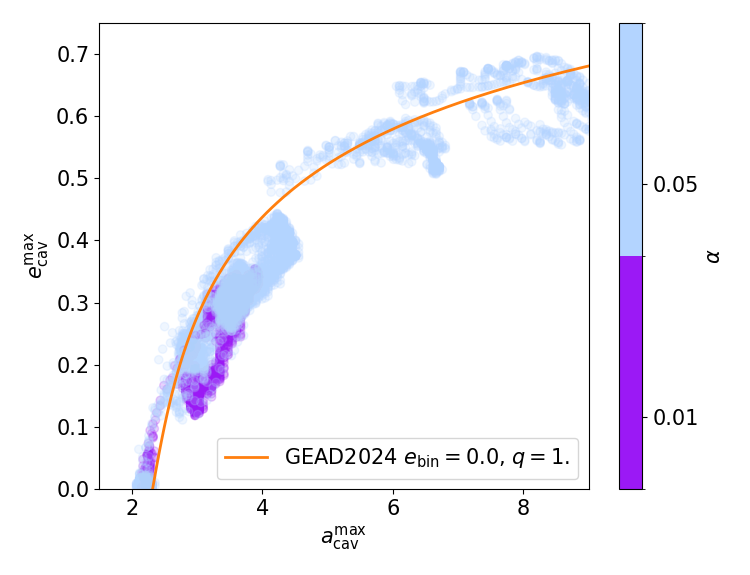}
   \caption{The relation $e_{\rm cav}^{\rm max}-a_{\rm cav}^{\rm max}$ for a selection of simulations from Groups 1, 3 and 4 with $e_{\rm bin}=0$, $q=1$ and different choices of $H/R$ and $\alpha$ (colors in the colorbar). Each data point represents a simulation snapshot. The orange solid line represent the $a^{\rm crit}$ orbital stability limit prescribed by \citet{georgakarakos2024}.}
\label{fig:acavecavq1}%
\end{figure}

In a gaseous disc, pressure gradients and viscosity attempt to push gas into the nominally unstable region. We identify these as ``hydrodynamic corrections'' to the gravitational stability limit. In Fig. \ref{fig:acavecavq1} we show how different values of $H/R$ and $\alpha$ affect the $e_{\rm cav}^{\rm max}$ vs. $a_{\rm cav}^{\rm max}$ relation, for $e_{\rm bin}=0.$ and $q=1$. The contribution of these fluid effects appears to be marginal: different values of $H/R$ and $\alpha$ does not affect the truncation $e_{\rm cav}^{\rm max}-a_{\rm cav}^{\rm max}$ relation itself, while they seem to affect the maximum value of $e_{\rm cav}^{\rm max}$ can achieve from the same initial conditions. The effect of hydrodynamic perturbations is quantified and discussed in more detail in Sec. \ref{sec:dependences}), while the long-term hydrodynamic evolution of the system is discussed at the end of Sec. \ref{sec:discussion}. 

\subsection{A semi-empirical formula for $a_{\rm cav}$}\label{sec:empirical}

\begin{table}
\centering
\caption{Values of $c_n$ coefficients in Eq. (\ref{eq:acav}) obtained from MCMC Bayesian inference on the full simulation dataset.  }\label{tab:MCMC}
\setlength{\tabcolsep}{3.5pt}
\begin{tabular}{ccccccccc|c}
\hline
  $c_1$ & $c_2$ & $c_3$ & $c_4$ & $c_5$ & $c_6$ & $c_7$ & $c_8$ & $c_9$ & $\sigma_0$ \\
\hline
  1.06 & 2.24 & 2.78 & -1.97 & 0.07 & 1.19 & -1.53 & -0.18 & 0.04 & 0.10 \\
\hline
\end{tabular}
\tablefoot{Notes: $\sigma_0$ is the standard deviation og the Gaussian likelihood (see Appendix \ref{appendix:MCMC}).
}
\end{table}

Inspired by this similarity between the extent of the cavity and that of the unstable region around the binary, and following a painstaking analysis of the dependence of the cavity size on the system parameters, we prescribe the cavity semi-major axis $a_{\rm cav}$ to be given by
\begin{equation}
    a_{\rm cav} =\frac{R_{\rm p}}{(1 - e_{\rm cav})},
\end{equation}
where we propose $R_{\rm p}$ to be the following semi-empirical formula describing the pericentre radius of the innermost stable orbit:
\begin{align}
R_{\rm p} &=
a_{\rm bin}\left[c_1 +
\underbrace{c_2 f_q }_{\text{Hill's radius}} +\underbrace{c_3 e_{\rm bin}+ c_4 e_{\rm bin}^2}_{\text{binary eccentricity}} + \right. \nonumber\\
&\quad \left. +
\underbrace{c_5 e_{\rm cav} +
c_6 e_{\rm cav}^2}_{\text{cavity eccentricity}} + \underbrace{c_7 e_{\rm bin} e_{\rm cav} \cos\left(\varpi_{\rm cav} - \varpi_{\rm bin}\right)}_{\text{phase coupling}} +\right.
\nonumber\\
&\quad \left. + 
\underbrace{c_8 \sqrt{\frac{\alpha}{\alpha_0}}}_{\text{viscosity term}} +
\underbrace{c_9 \frac{h}{h_0}}_{\text{aspect ratio term}}\right].\label{eq:acav}
\end{align}
where $f_{\rm q}$ is defined as
\begin{align}
    f_{q}=\begin{cases}
(q/3)^{1/3}, & \text{if } q<q_{\rm cut} \\
(q_{\rm cut}/3)^{1/3},  & \text{if } q\geq q_{\rm cut}
\end{cases},
\end{align}
with $q_{\rm cut}=0.4$, $\alpha_0=0.1$ and $h_0=0.1$.
Note that $a_{\rm bin}f_q=R_{\rm Hill}$ corresponds to the Hill's radius. The coefficients $c_1\textrm{--}\,c_9$ are dimensionless parameters that we constrain using Monte Carlo Markov Chain (MCMC) Bayesian inference. We discuss the role of each term in Sec. \ref{sec:dependences}. Details about the MCMC procedure are reported in Appendix \ref{appendix:MCMC}, while the inferred values for $c_1\textrm{--}c_9$ are reported in Tab. \ref{tab:MCMC}.

The expression in Eq. (\ref{eq:acav}) sums multiple physically motivated contributions that either increase or decrease the extent of the unstable region around the binary.
Intuitively, each term has a geometrical interpretation that shifts the effective gravitational influence of the binary outward or inward. The term $c_1\sim 1$ sets the reference length of the problem, $a_{\rm bin}$. The coefficient $c_2$ is proportional to the size of the Hill sphere of the secondary mass for mass ratios $q<0.4$, and remains constant  for larger mass ratios\footnote{Sticking to the geometrical interpretation, we speculate that this effect arises because, as the mass ratio increases, the centre of mass of the binary moves towards the secondary, reducing the effective reference length to the maximum distance of the secondary from the centre of mass of the binary, namely $a_2 = a_{\rm bin}/(1+q)$. This inward shift compensates for the increase in the Hill radius.}.

The coefficients $c_3$ and $c_4$ account for the binary eccentricity, which influences the location of the separation of the binary at apocentre (i.e., it marks the closest point to the cavity edge). Coefficients $c_5$ and $c_6$ account for the cavity eccentricity, as higher eccentricity brings the cavity pericentre closer to the binary apocentre. The value of $c_5=0.07\ll c_6$ suggests that the dependence on $e_{\rm cav}$ in $R_{\rm p}$ is mainly quadratic. Coefficient $c_7<0$ captures the reduced stability of disc orbits when $\varpi_{\rm cav}-\varpi_{\rm bin}\approx {\rm \pi}$, that is when the cavity pericentre aligns with the binary apocentre, thereby increasing the pericentre radius $R_{\rm p}$ of the last stable orbit. Finally, coefficients $c_8\sim -0.18$ and $c_9\sim 0.04$, describe the mild dependence of the innermost stable orbit on viscosity $\alpha$ and disc aspect ratio $H/R$, with contributions to the value of $R_{\rm p}$ of $0.04\lesssim \Delta R_{\rm p}(\alpha)/a_{\rm bin}\leq 0.18$ and $0.01\lesssim \Delta R_{\rm p}(H/R)/a_{\rm bin}\leq 0.04$, respectively.

\begin{figure}
   \centering   \includegraphics[width=\columnwidth]{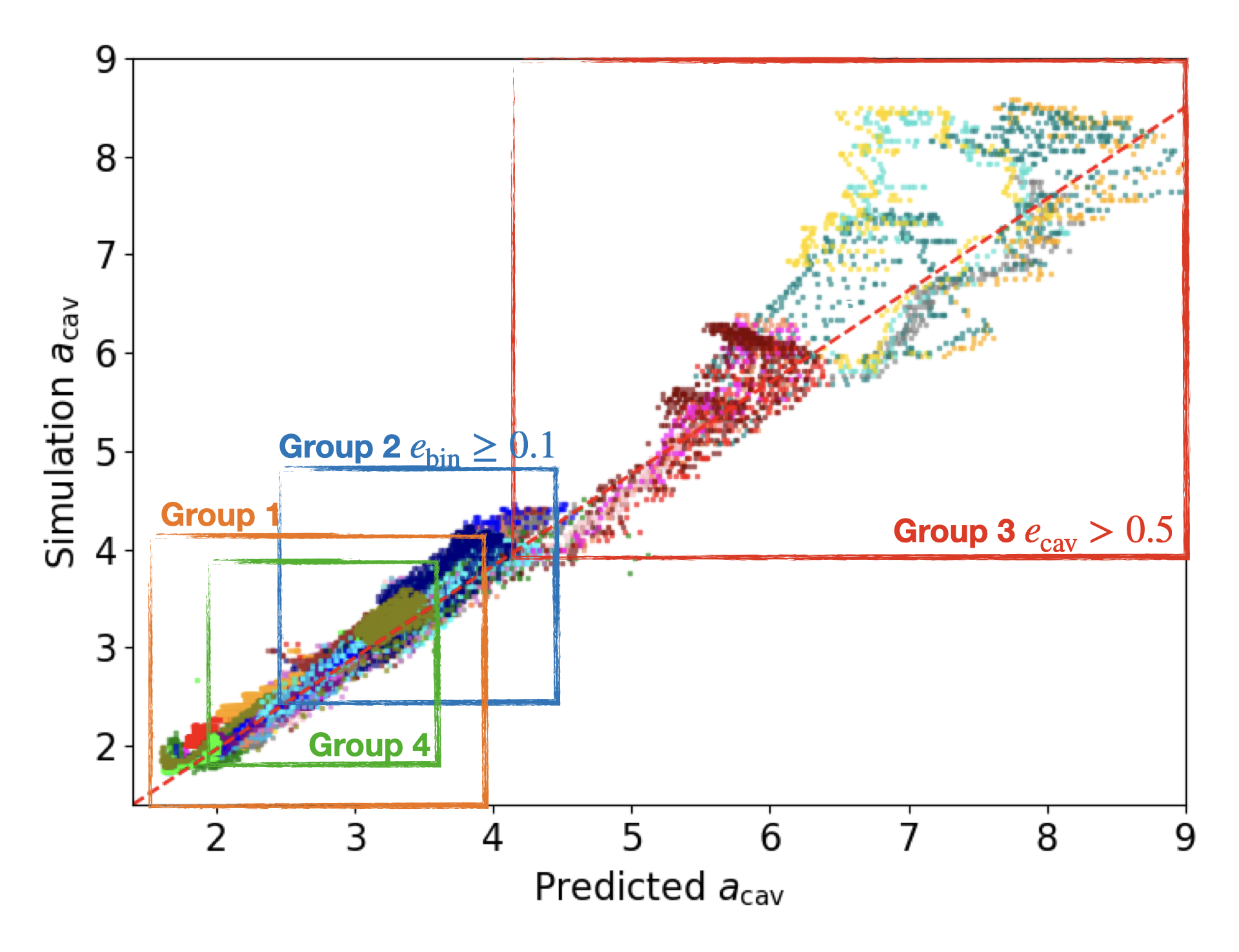}\\
   \includegraphics[width=\columnwidth]{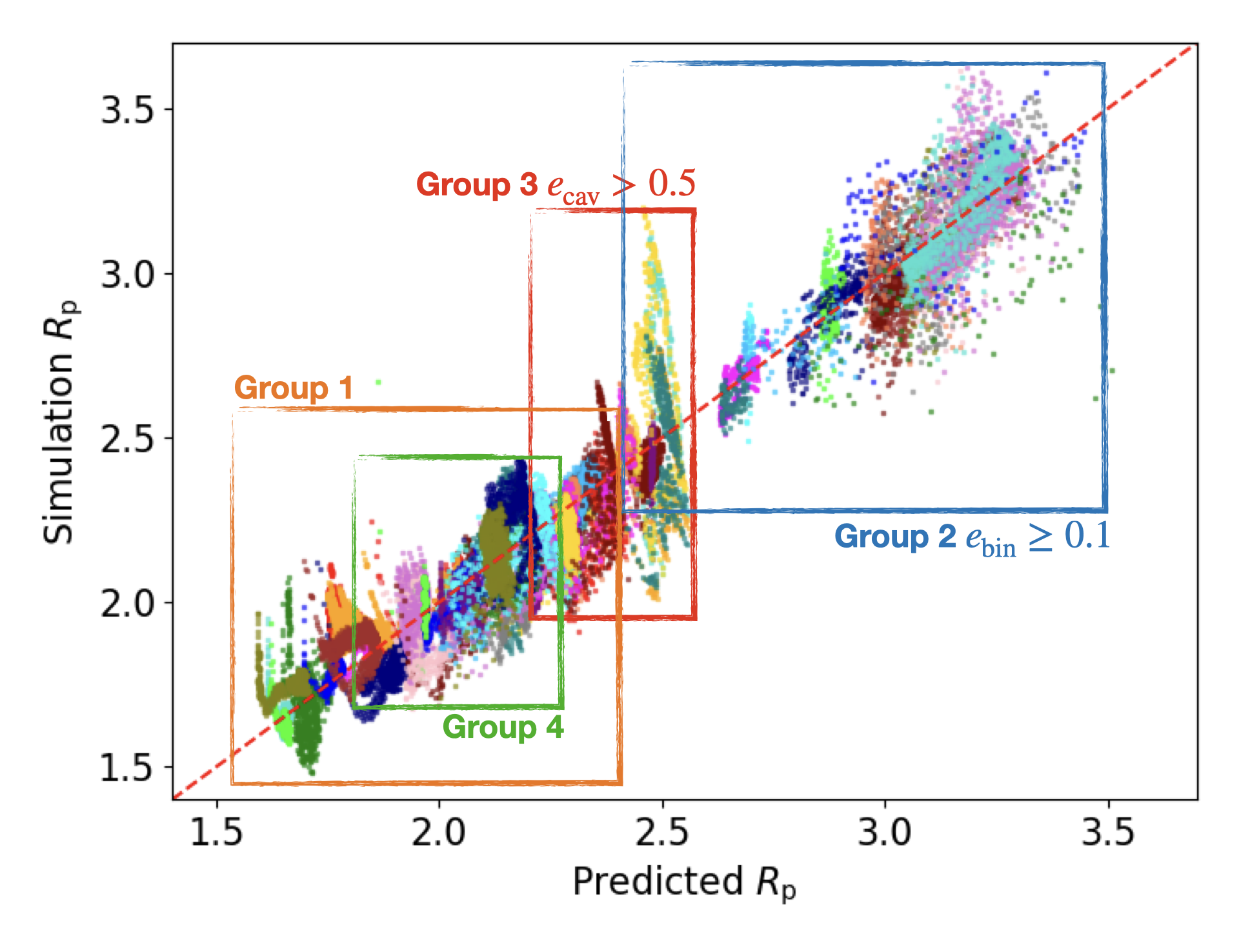}
   \caption{Summary plot of simulation values of $a_{\rm cav}$ (top panel) and $R_{\rm p}$ (bottom panel) compared to theoretical predictions from (\ref{eq:acav}). Each point represent a snapshot of a simulation, each color represents a different numerical simulation. Colored rectangles indicate regions corresponding to different Groups. Orange and green rectangles highlight the extent of all simulations from Group 1 and Group 4, respectively. Blue and red rectangles highlight simulations with $e_{\rm bin}\geq 0.1$ and $e_{\rm cav}>0.5$ from Group 2 and Group 3, respectively.}
\label{fig:acavvsatheor}%
\end{figure}

\begin{figure*}
   \centering   
   \includegraphics[width=\columnwidth]{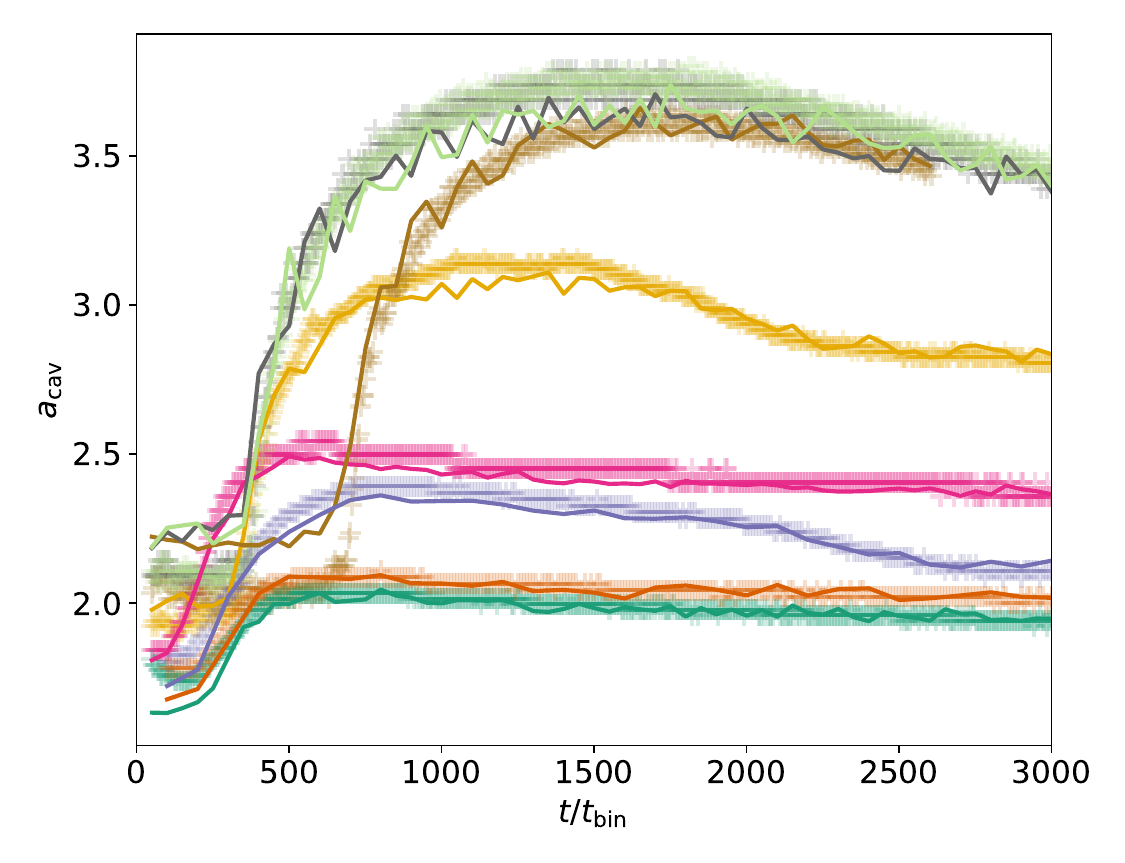}\includegraphics[width=\columnwidth]{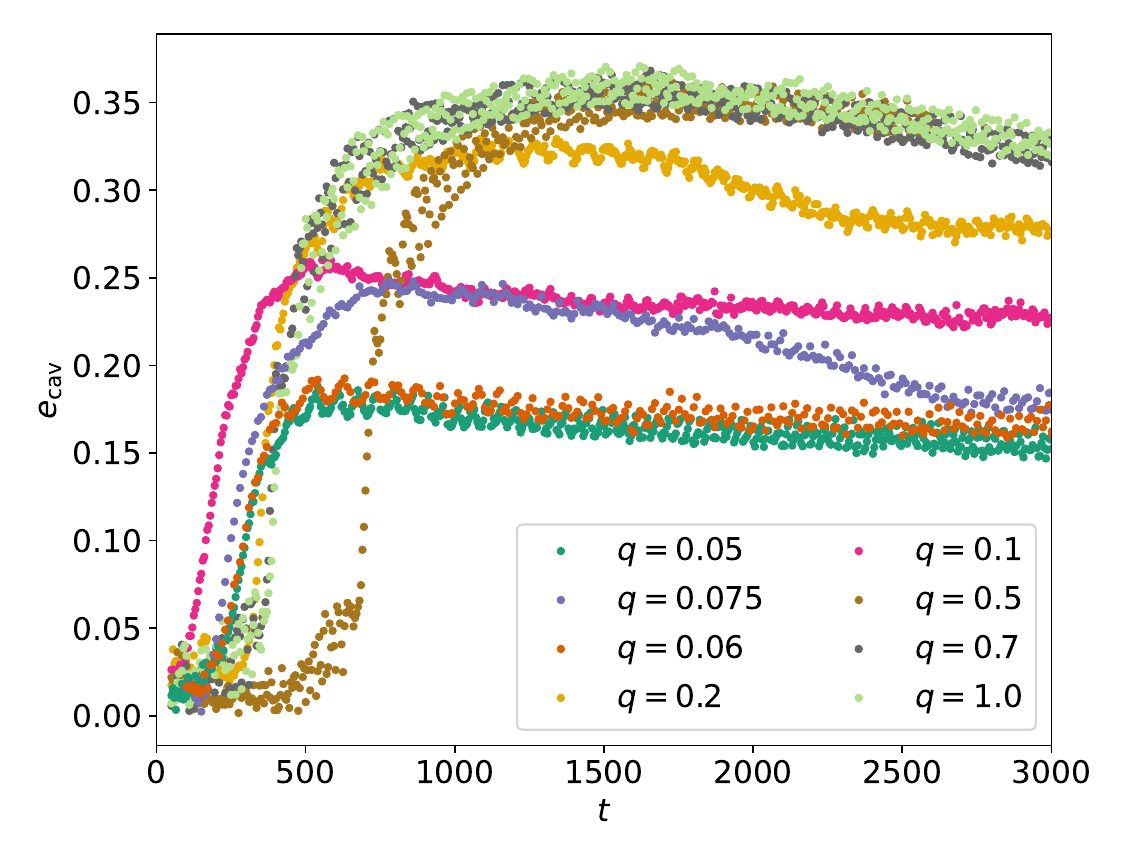}
   \caption{Evolution of cavity semi-major axis $a_{\rm cav}$ and cavity eccentricity with time. Left panel: time evolution of the cavity semi-major axis $a_{\rm cav}$, for various binary mass ratios $q=\{0.05, 0.06, 0.075, 0.1, 0.2, 0.5, 0.7, 1.0\}$ (for fixed $e_{\rm bin}=0$ and $H/R=0.05$, i.e. a selection of simulations from Group 1), color legend in the right panel. Crosses correspond to individual simulation snapshots, solid lines are obtained using Eq. (\ref{eq:acav}). Right panel: same as left panel but showing the time evolution of $e_{\rm cav}$. Each point corresponds to individual simulation snapshots.}
\label{fig:qdep}%
\end{figure*}

\begin{figure*}
   \centering  
\includegraphics[width=\columnwidth]{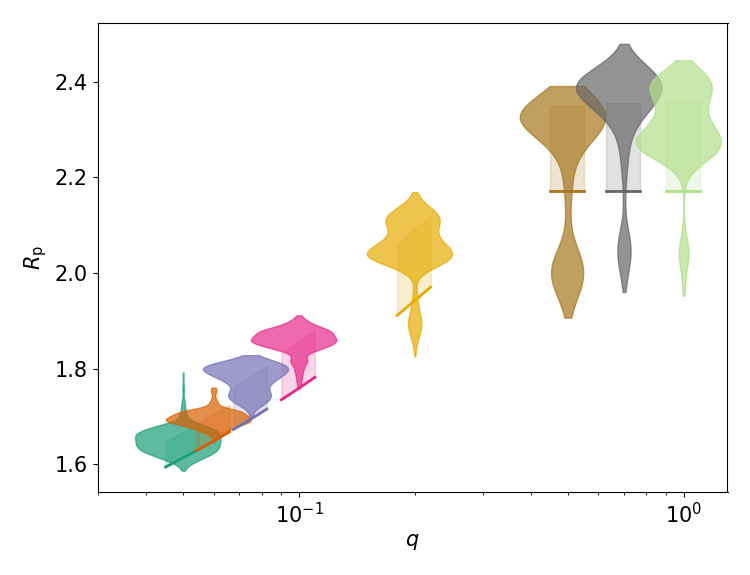}\includegraphics[width=\columnwidth]{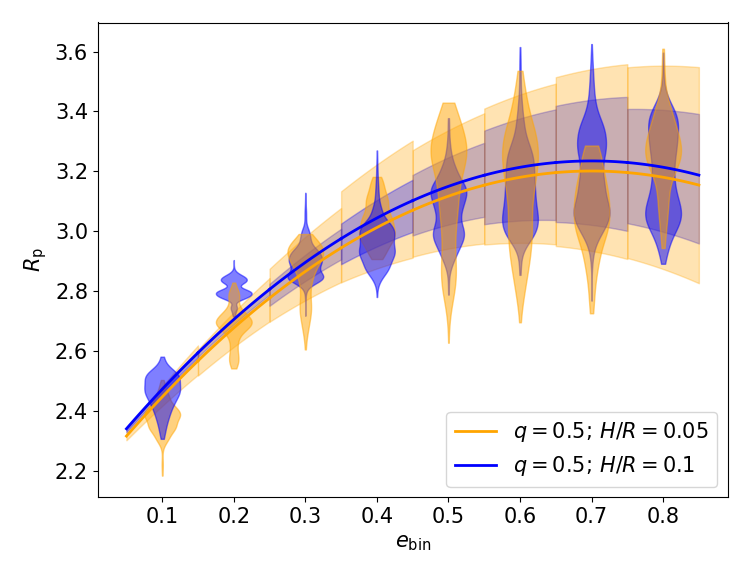}

   \caption{Dependence of cavity pericentre radius $R_{\rm p}$ on the system parameters. Left panel: $R_{\rm p}$ as a function of binary mass ratio $q=\{0.05, 0.06, 0.075, 0.1, 0.2, 0.5, 0.7, 1.0\}$ (for fixed $e_{\rm bin}=0$ and $H/R=0.05$, i.e. a selection of simulations from Group 1). Violin plots track the range of $R_{\rm p}$ explored throughout the simulations, while shaded regions represent the predicted range of $R_{\rm p}$ from Eq. (\ref{eq:acav}) using the maximum disc eccentricity $e_{\rm cav}$ reached during the simulation. Solid lines show the prediction for $R_{\rm p}$ assuming $e_{\rm cav}=0$. Right panel: same as left panel but showing $R_{\rm p}$ as a function of binary eccentricity $e_{\rm bin}=\{0.1, 0.2, 0.3, 0.4, 0.5, 0.6, 0.7, 0.8\}$ and disc aspect ratio $H/R=\{0.05,0.1\}$ (for fixed $q=0.5$ and  $\alpha=0.005$, i.e. a selection of simulations from Group 2).}
\label{fig:edep}%
\end{figure*}

\begin{figure}
   \centering  
   \includegraphics[width=\columnwidth]{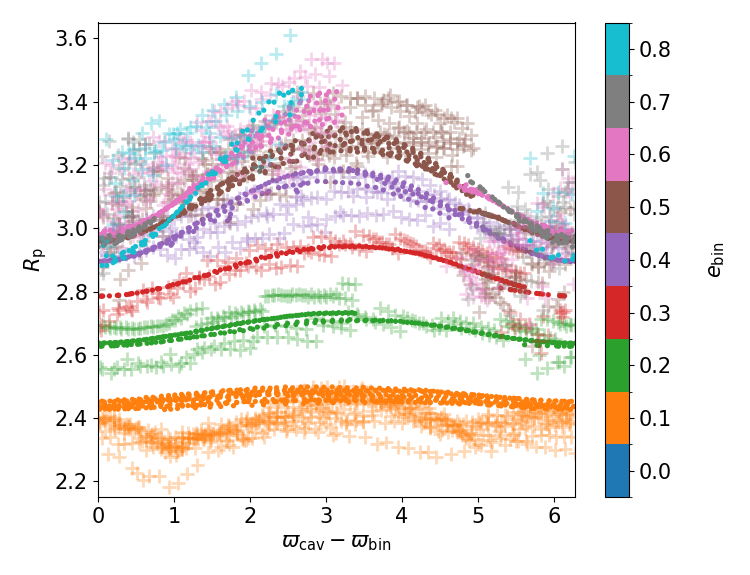}
   \caption{Dependence of $R_{\rm p}$ as a function of relative orientation of disc and binary pericentre longitudes, $\varpi_{\rm cav}-\varpi_{\rm bin}$, as a function of binary eccentricity $e_{\rm bin}$, for simulations with $e_{\rm bin}=\{0.1, 0.2, 0.3, 0.4, 0.5, 0.6, 0.7, 0.8\}$,  $H/R=0.05$, $\alpha=0.005$ (i.e. a selection of simulations from Group 2, shown in right panel of Fig. \ref{fig:edep}).  
   Crosses represent individual simulation snapshots, and squares show predictions from Eq. (\ref{eq:acav}); colours indicate different $e_{\rm bin}$ values as reported in the colourbar. Note that some simulations complete multiple precession cycles with different values of $e_{\rm cav}$, so a given value of $\varpi_{\rm cav}-\varpi_{\rm bin}$ can correspond to multiple values of $a_{\rm cav}$.}
\label{fig:omdep}%
\end{figure}

We show in Fig. \ref{fig:acavvsatheor} a comparison between the theoretical prediction in Eq. \ref{eq:acav} and the value extracted from the numerical simulation. Figure \ref{fig:qdep} shows  the time evolution of $a_{\rm cav}$ and $e_{\rm cav}$ as a function of the binary mass ratio $q$. Figure \ref{fig:edep} shows violin plots of $R_{\rm p}$ as a function of binary mass ratio $q$ and eccentricity $e_{\rm bin}$. Violin plots highlight the variability of $a_{\rm cav}$ throughout the duration of the simulations. Theoretical predictions from Eq. (\ref{eq:acav}) are shown as shaded regions in left panels and solid curves in right panels. Finally, Fig. \ref{fig:omdep} shows the dependence of $R_{\rm p}$ on the relative orientation of binary and cavity pericentres $\varpi_{\rm cav}-\varpi_{\rm bin}$.

\subsection{Dependence on $H/R$ and $\alpha$  }\label{sec:dependences}

\begin{figure}
\includegraphics[width=\columnwidth]{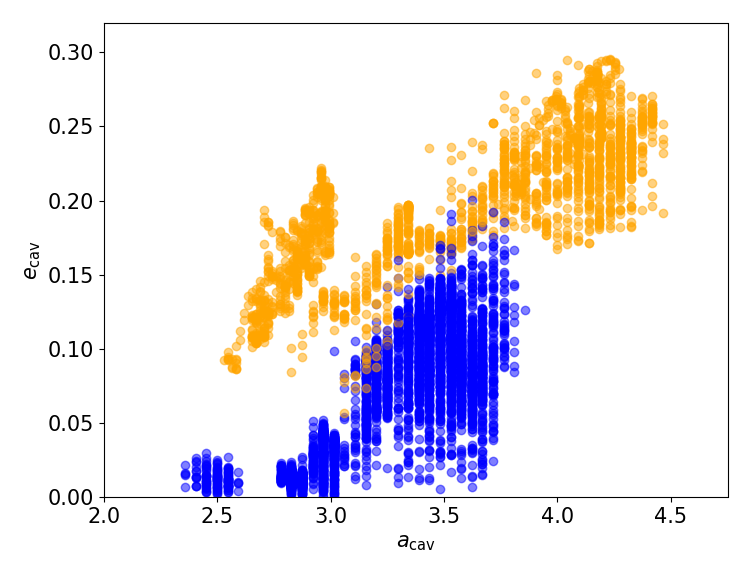}
\includegraphics[width=\columnwidth]{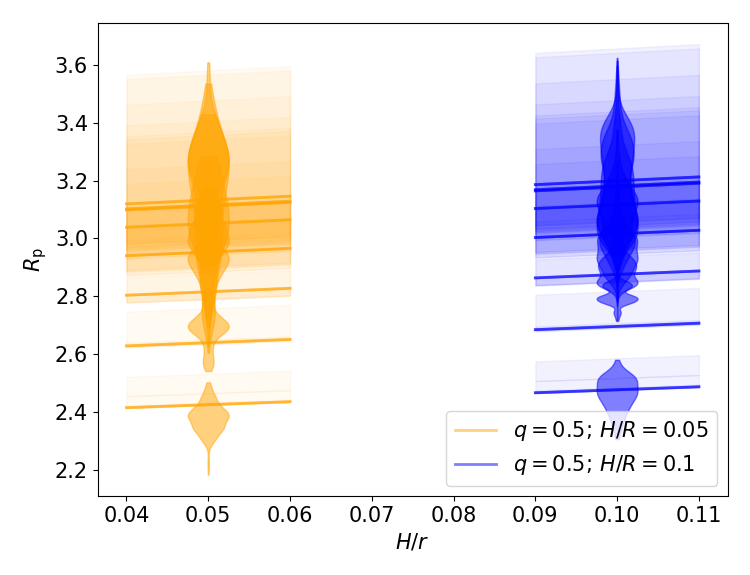}
   \caption{Dependence of $e_{\rm cav}$, $a_{\rm cav}$, $R_{\rm p}$ on $H/R$. Top panel: $a_{\rm cav}$ as a function of $e_{\rm cav}$ (i.e. as Fig. \ref{fig:dataset}) restricted to Group 2 simulations with $q=0.5$, varying the whole range of $e_{\rm bin}$ (as in Fig. \ref{fig:edep}) and $H/R=\{0.05,0.1\}$. Orange dots indicate simulations with $H/R=0.05$, blue points simulations with $H/R=0.1$. For fixed $q$, $e_{\rm bin}$ and $e_{\rm cav}$, larger values of $H/R$ correspond to larger $a_{\rm cav}$, but, most importantly, since they reach lower cavity eccentricities, they seem smaller. Bottom panel: Violin plots showing values of $R_{\rm p}$ during the simulations as a function of $H/R$. Shaded areas mark the predictions by Eq. \ref{eq:acav}, solid lines the predictions for $e_{\rm cav}=0$. Larger values of $H/R$ counterintuitively produce slightly larger values of $R_{\rm p}$, as predicted by Eq. (\ref{eq:acav}).}
\label{fig:H}%
\end{figure}

While larger values of $\alpha$ lead to smaller innermost stable orbits ($c_8<0$), consistent with  previous studies, the small, but positive, value for $c_9=0.04$ indicates that larger values of $H/R$ pushes the innermost orbit outward, contrary to the expectation that thicker discs produce smaller cavities. We speculate that this effect arises because orbital stability (independent of hydrodynamics) sets the location of the innermost orbit, whereas the surface density maximum is displaced outward by approximately $H$, which defines the reference length scale of stable density gradients in accretion discs. This result, while seemingly at odds with the past literature, where thicker discs are typically associated with smaller cavities, is actually consistent with the previous findings: disc eccentricity saturates to lower values in thicker discs, resulting ultimately in a smaller cavity. This behaviour is illustrated in Fig. \ref{fig:H}, where simulations in Group 2 show that  $H/R=0.05$ discs generally have larger and more eccentric cavities than the corresponding $H/R=0.1$ cases; nevertheless, for larger $H/R$ truncation is mildly more effective ($c_9\gtrsim0$), producing slightly larger $R_{\rm p}$ values than for $H/R=0.05$.

\section{Discussion}\label{sec:discussion}

\begin{figure}
   \centering   \includegraphics[width=\columnwidth]{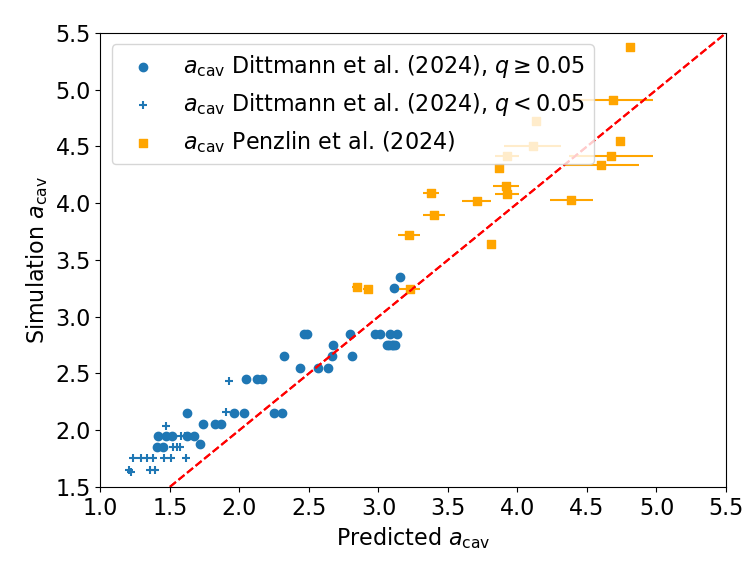}
   \caption{Comparison with other works. Same as top panel of Fig. \ref{fig:acavvsatheor} but using simulations from \citet{dittman2024} (blue circles and crosses) and \citet{penzlin2024} (orange squares). Horizontal lines mark the range between minimum and maximum predicted $a_{\rm cav}$ (using Eq. \ref{eq:acav}) due to the unknown value of $\varpi_{\rm cav}-\varpi_{\rm bin}$. Crosses are simulations using mass ratios $q<0.05$ while circles are for $q\geq 0.05$ in the \citet{dittman2024} sample. Data points not showing the horizontal bar represent simulations with circular binaries, since Eq. (\ref{eq:acav}) predict no variability with $\varpi_{\rm cav}$ in the cavity size for $e_{\rm bin}=0$. Nevertheless, we note that some variability, not captured by our model, can be observed in our circular binary simulations (see discussion in Sec. \ref{sec:uncertainties}).}
\label{fig:penzlindittman}%
\end{figure}

\subsection{The nature of tidal truncation}

The results presented highlight that, for mass ratios $q\geq 0.05$, the mechanism of tidal truncation shares some important similarities with the dynamical processes that describe the stability of the orbits in the restricted three-body problem. In this framework, the pericentre distance is determined by celestial mechanical considerations, with mild corrections from hydrodynamic effects such as viscosity and pressure and acts on short, timescales, of the order of a few orbital times, as short-term processes. At the same time, viscosity and pressure, coupled with the action of individual resonances that gain strength within the disc, act on longer-timescales as long-term processes. More work is required to define whether tidal truncation consists of an interplay between short-term and long-term processes, or whether they operate independently as mutually exclusive mechanisms. The small contributions from coefficients $c_8$ and $c_9$ in Eq. (\ref{eq:acav}) appear to suggest that, in our simulations, viscosity and disc thickness do not directly control truncation (which depends only weakly on them); rather, they indirectly yield smaller cavities by limiting the saturated eccentricity, an effect that could otherwise be misinterpreted as less efficient truncation. We cannot exclude that longer-term effects, that are not probed by our numerical setups, might lead to a stronger dependence on these parameters at later times.

In general, independently of the mechanism in action, cavity eccentricity appears to be a key element to determine the cavity size. However, its saturation level is known to be tightly linked to the thermodynamical treatment of the disc, including heating and cooling processes \citep{sudarshan2022,penzlin2025}. 
Moreover, we note that the assumption of a locally isothermal equation of state causes the angular momentum flux \citep{miranda2020} and angular momentum deficit \citep[AMD, ][]{teyssandier2016} to not be conserved. The flow instead conserves equivalent sound-speed weighted quantities. Since the disc eccentricity evolution is tied to the conservation of the AMD, the choice of a locally isothermal profile may play a role in shaping the specific eccentricity ``history'' of the cavity. Properly capturing these thermodynamic effects is therefore crucial for understanding the long-term evolution of cavity eccentricity, and thus the resulting disc truncation and structure.

When it is possible to reach a quasi-steady state, this is achieved through the balance between viscous stresses, gravitational forces and other angular momentum transport processes in the disc (e.g. transport and deposition through density waves \citealp{rafikov2016}, or disc winds \citealp{bai2016}) setting a constant angular momentum flux through the disc (e.g. \citealp{munoz2018}). The "balance" is global: the viscous torques, angular momentum transport through spiral density waves in the outer disc, and gravitational torque causes the material to be supplied to the cavity edge at a constant rate, and the dynamical chaos within the cavity funnels it into the sinks at the same rate. Indenpendently of whether orbital instability or other processes are dominant in shaping the cavity size, the unstable region represents a dynamical transition zone similar to the Innermost Stable Circular Orbit (ISCO) in black hole accretion discs.
In this picture, within the unstable zone act an open boundary. Once material reaches its edge, it enters the chaotic zone where stable circular orbits no longer exist. At this point, the gas has two options \citep{tiede2022}: i) the material falls onto the sinks (or their circumstellar discs), which act as the ultimate sinks for both mass and angular momentum; ii) the material is thrown against the other edge of the cavity (in 3D simulations the material can also fly above the cavity wall and be ejected from the system).  This is inherently  consistent with the fact that circumbinary discs are known to be actively accreting systems.

We finally emphasize that when orbital instability is in action to define the cavity boundary, the binary continues to inject angular momentum and energy into the disc, while conversely experiencing a back-reaction torque through its coupling with density perturbations. Beyond the resonance overlap region, we expect individual resonances to become stronger as the disc eccentricity evolves, producing a slow long-term evolution towards the final disc geometric configuration, and, when possible, the reach of a quasi-steady state accretion flow.

\subsection{Discrepancies and uncertainties}\label{sec:uncertainties}

In general, from Fig. \ref{fig:acavvsatheor} we note that the prescription in Eq. (\ref{eq:acav}) matches $a_{\rm cav}$ and $R_{\rm p}$ from the simulations with an uncertainty of $\lesssim 15\%$. 
This uncertainty increases for the largest cavities, that are associated with values of $e_{\rm cav}\gtrsim 0.5$, in particular: simulations  05A6$e_{\rm d}$, 05B6$e_{\rm d}$, 1A6$e_{\rm d}$,
1B6$e_{\rm d}$.
In these cases, cavities are mainly larger than predicted by Eq. (\ref{eq:acav}), so that for these simulations it acts as a lower limit; this trend is visible in Fig. \ref{fig:acavvsatheor} where the prescription tends to underestimate cavity sizes.

In this context, it is important to distinguish between ``systematic discrepancy'' (persistent deviations from the expected value) and ``variability'' (fluctuations around a mean value). 
The ``lower-limit'' behaviour for $e_{\rm cav}\gtrsim 0.5$ is an example of a ``systematic discrepancy'' which, we speculate, has two possible interpretations. First, these discs are characterised by strong dynamical perturbations (see central panel of bottom row in Fig. \ref{fig:simexample}): tidal streams stripped from one edge of the cavity collide with the opposite side (e.g., \citealp{shi2012,farris2014}), while steep eccentricity and pericentre longitude gradients can lead to shocks due to intersecting orbits \citep{ogilvie2001,statler2001}. These effects potentially deplete the material near the cavity edge, increasing the measured cavity size. In fact, Eq. (\ref{eq:acav}) marks the innermost radius beyond which orbits remain stable, but it does not prevent other concurrent mechanisms from truncating or depleting the disc at larger radii. Second, the functional form of Eq. (\ref{eq:acav}), where the factor $a_{\rm cav}\propto (1-e_{\rm cav})^{-1}$ dominates the $e_{\rm cav}$ dependence in Eq. (\ref{eq:acav}), becomes inadequate for describing large values $e_{\rm cav}$.  

Regarding ``variability'', secular and resonant effects produce oscillations of $a_{\rm cav}$ around a mean value. 
The full dynamics of circumbinary discs involves coupled secular and resonant oscillations of $a_{\rm cav}$, $e_{\rm cav}$ and $\varpi_{\rm cav}$, that are not captured by Eq. (\ref{eq:acav}). 
These appear as phased or antiphased periodic oscillation between both $a_{\rm cav}$ and $e_{\rm cav}$ across a median value (resonant) or oscillations of $e_{\rm cav}$ only at nearly constant $a_{\rm cav}$ (secular). In this framework, the innermost stable orbit sets a median theoretical cavity size $a_{\rm cav}^{\rm th}$, with the simulations exploring a range of values around this median.

Theoretical and numerical studies about orbital stability indicate that stable regions in the $e\textrm{--}a$ plane are inherently ``ragged'', due to chaotic behaviour emerging where resonances overlap (e.g., see Fig. 3 of \citealt{shevchenko2015} or Fig. 6 in \citealt{adelbert2023}). This produces abrupt transitions between adjacent stable and unstable regions, causing variations in $a_{\rm cav}$ that cannot be captured by a smooth prescription such as Eq. (\ref{eq:acav}). 

Finally, to assess the reliability of our results, we compare the predictions from Eq. (\ref{eq:acav}) with the simulations from two different works \citep{penzlin2024,dittman2024} in Fig. \ref{fig:penzlindittman}. Despite differences in the approach used to measure the cavity size\footnote{The location of the $10\%$ from the maximum value of the surface density profile in \citet{penzlin2024} and the location of $\min_a(|\partial^2\Sigma/\partial a^2|)$ in \citet{dittman2024}.} (\citealp{dittman2024}, also have binaries with $q<0.05$ in their sample), the results in these works show remarkable agreement with Eq. (\ref{eq:acav}). 

\section{Conclusion}\label{sec:conclusion}

In this work we have presented a suite of 80 numerical simulations and revisited the long-standing problem of tidal truncation in circumbinary discs, focusing in particular on the role played by disc eccentricity in setting the size and geometry of the central cavity. By analysing the extensive dataset of three-dimensional SPH simulations using the code \textsc{phantom} \citep{price2018a}, spanning a wide range of binary and disc parameters, we show that:

(i) The growth of disc eccentricity and the expansion of the cavity occur concurrently and on comparable dynamical timescales, to the extent they can be considered tightly correlated. The often-reported slow evolution of the cavity size over hundreds or thousands of binary orbits needs to be interpreted considering this dependence on the cavity eccentricity.

(ii) The cavity size $a_{\rm cav}$ follows trends that strongly resemble those of chaotic orbit boundaries derived from purely gravitational considerations. Consistently with this insight, we introduced in Eq. (\ref{eq:acav}) a semi-empirical formula that predicts the cavity size as a function of the instantaneous dynamical properties of the system. The prescription shows remarkable agreement within our own sample (Fig. \ref{fig:acavvsatheor}) and other works (Fig. \ref{fig:penzlindittman}).

(iii) The impact of viscosity and pressure on the truncation radius is sub-dominant compared to that of the binary orbital parameters and disc eccentricity. The apparent dependence of cavity size on disc thickness reported in previous studies can be largely understood as an indirect effect: thicker, viscous discs tend to saturate at lower eccentricities, which in turn leads to smaller cavities.

(iv) Circumbinary discs should be regarded as fundamentally eccentric objects, whose global eccentric modes play an active role in shaping the disc structure. Treating eccentricity as a passive by-product of binary-disc interaction misses essential aspects of the dynamics.

(v)  We speculate that processes analogous to orbital stability in the restricted three-body problem participate in tidal truncation acting on short-timescales. Whereas, on longer-timescales, the resonant-viscous torque balance from individual resonances is altered by the growth of disc eccentricity, impacting on the long-term evolution of the system. Our results do not allow us to determine whether these mechanisms undergo a complex interplay or whether they act as separate processes. Future work focusing on the long-term evolution of the system remains essential to fully disentangle their respective effects.

This paper is the first in a series devoted to understanding the dynamics of eccentric circumbinary discs. In forthcoming work, we will explore in more detail the excitation, saturation, and long-term evolution of eccentric eigenmodes, their coupling to binary orbital evolution, and the consequences for mass accretion, variability, and disc–binary angular momentum exchange. Together, these studies aim to build a coherent dynamical picture of circumbinary discs in which eccentricity is a central, predictive ingredient.

\begin{acknowledgements}
The authors thank the anonymous referee for their comments that improved the manuscript. E.R. thanks Giuseppe Lodato, Daniel Price, Giovanni Rosotti, Nikolaos Georgakarakos, Alessia Franchini and Guglielmo Mastroserio for useful discussions. E.R. thanks Callum Fairbairn, Gordon Ogilvie, and Roman Rafikov for their comments on a preliminary version of the left panel of Fig. \ref{fig:dataset} during a visit to DAMTP (Cambridge) in November 2022.
E.R. acknowledges financial support from the European Union's Horizon Europe research and innovation programme under the Marie Sk\l{}odowska-Curie grant agreement No. 101102964 (ORBIT-D).
ER, EL and GL acknowledge financial support from the European Research Council (ERC) under the European Union’s Horizon 2020 research and innovation programme (grant agreement No. 864965, PODCAST).
ER and RA acknowledge financial support from the European
Research Council (ERC) under the European Union’s Horizon 2020
research and innovation programme (grant agreement no. 681601, BuildingPlanS).
ER also acknowledges support from the European Union (ERC Starting Grant DiscEvol, project number 101039651). 
This project has received funding from the European Union's Horizon 2020 research and innovation programme under the Marie Sk\l{}odowska-Curie grant agreement No 823823 (DUSTBUSTERS).
EL acknowledges funding from the Japan Society for the Promotion of Science through a JSPS International Research Fellowship.
RA acknowledges funding from the Science \& Technology Facilities Council (STFC) through Consolidated Grant ST/W000857/1.
Views and opinions expressed are, however, those of the author(s) only and do not necessarily reflect those of the European Union or the European Research Council.
The simulations performed for this paper used the DiRAC Data Intensive service at Leicester, operated by the University of Leicester IT Services, which forms part of the STFC DiRAC HPC Facility (www.dirac.ac.uk).
Fig. \ref{fig:simexample} was created using \textsc{splash} \citep{price07a}. All the other figures were created using \textsc{matplotlib} python library \citep{hunter2007}.
\end{acknowledgements}

\bibliographystyle{aa}
\bibliography{biblio}

\appendix

\section{Initialising an eccentric disc}

\subsection{Disc setup}\label{appendix:eccentricdiscs}

In this Appendix we discuss how to initialise SPH particles to set up a Keplerian eccentric disc with eccentricity profile $e(a)$, longitude of pericentre profile $\varpi(a)$ and circular surface density profile $\Sigma(a)$ (see Eq. (\ref{eq:sigma}) in Sec. \ref{sec:analysis}). 
The geometry and local surface density distribution $\Sigma_{\rm loc}(a,\phi)$ of the system are fully determined by these three quantities, and can be written as: 

\begin{equation}
\Sigma_{\rm loc}(a,\phi)=\Sigma(a)\frac{a\Omega_0(a)}{J(a,\phi)\Omega(a,\phi)},\label{eq:sigmaapp}
\end{equation}
where $J(a,\phi)$ is the determinant of the Jacobian matrix of the coordinate transform $(x,y)\rightarrow (a,\phi)$, $\Omega_0(a)$ is the orbital frequency for an orbit with semi-major axis $a$ and $\Omega(a,\phi)$ is the angular velocity along the orbit \citep{ragusa2024}. 

In \textsc{phantom}, for circular discs, the particles are initially distributed using an accept-reject approach that distributes the particles with the desired radial profile. To create the initial distribution of particles for an eccentric disc, we use a similar approach and proceed as follows to obtain the correct $(a,\phi)$ distribution (we assume $d\varpi/da=0$):

\begin{enumerate}
\item We randomly generate a value of $a$ from a uniform distribution between inner and outer edges of the disc.\\

\item We randomly generate a value of $\phi$. To do so, we first generate from a uniform distribution a mean anomaly $M$, we convert it to eccentric anomaly $E$ solving the equation $M=E-e\sin(E)$ using the Newton-Raphson method, and finally we convert $E$ to true anomaly 
\begin{equation}
f=2\,\arctan\left[\sqrt{\frac{1+e}{1-e}} \,\tan\left(\frac{E}{2}\right)\right],
\end{equation}
and consequently set $\phi=f$. This produces the correct azimuthal distribution.\\

\item We generate randomly from a uniform distribution $[0,\max(\Sigma(a))]$ a final variable $p$  to accept or reject the couple $(a,\phi)$ we generated. If the value of $p$ satisfies 
\begin{equation}
p<2{\rm \pi}\,\Sigma(a)\,a. \label{eq:pcrit}
\end{equation}
the couple $(a,\phi)$ is accepted.

This criterion for acceptance is obtained by noting that the flux
\begin{equation}
 \mathcal F(a)= J(a,\phi) \Omega(a,\phi) \Sigma_{\rm loc}(a,\phi)=\Sigma(a) a\Omega_0\label{eq:flux}
\end{equation} 
is constant along the orbit. Combining the expression for the mass contained within a certain radius ${\rm d}M=\int\Sigma J{\rm d}a{\rm d}\phi$ with Eq. (\ref{eq:sigmaapp}) and (\ref{eq:flux}). Integrating across the whole orbit, we obtain that the amount of mass to be found within an annulus of thickness ${\rm d}a$ is 
\begin{equation}
    \frac{{\rm d}M}{{\rm d}a}=2{\rm \pi}\Sigma(a)a,
\end{equation}
from which Eq. (\ref{eq:pcrit}) is obtained.\\
\item Coordinate $z$ is sampled randomly from a Gaussian distribution with $\sigma=H$. \\

\item Particle coordinates are then set using 
\begin{align}
x(a,\phi)&=R(a,\phi)\cos[\phi],\\
y(a,\phi)&=R(a,\phi)\sin[\phi],\\
z&=z,
\end{align}
where $R$ is
\begin{equation}
R(a,\phi)=\frac{a[1-e(a)^2]}{1+e(a)\cos[\phi-\varpi(a)]}.
\end{equation}

\item Particle velocities are set assuming Keplerian orbits with no pressure corrections 
\begin{align}
v_{R}(a,\phi)&=   a\Omega_0\frac{e(a)}{\sqrt{1-e^2(a)}}\sin[\phi-\varpi(a)],\\
v_{\phi,{\rm K}}(a,\phi)&=    a\Omega_0\frac{1+e(a) \cos[\phi-\varpi(a)]}{\sqrt{1-e^2(a)}}\\
v_z&= 0,
\end{align}
where $\Omega_0=\sqrt{GM/a^3}$ is the orbital frequency.
\end{enumerate}

In general this implementation neglects the different vertical displacement that is expected to occur at apocentre and pericentre (e.g. \citealp{ragusa2024}). Similarly, it neglects the vertical velocity resulting from the change of the vertical equilibrium along the orbit and the pressure corrections to the azimuthal velocity. We note this implementation is not supposed to represent the perfect equilibrium state of an eccentric disc, but a reasonably stable starting point from which the simulation can evolve to its most stable structure.

\subsection{Setting up eccentric eigenmodes}\label{appendix:ecceigenmode}

To obtain the nonlinear eccentric eigenmodes used to generate the initial conditions for the Group 4 simulations we solve Equation (51) of \citet{ogilvielynch2019} as a nonlinear eigenvalue problem for the precession frequency using a shooting method\footnote{A version of the code used to solve the eigenvalue problem and generate the orbital elements and fluid properties can be found at: \url{https://github.com/ElliotLynch/EccentricDiscSolvers.git}.}.  
We assume the surface density profile to be
\begin{equation}
\Sigma(a)=\Sigma_0\left( \frac{a}{a_{\rm in}} \right)^{-p}\left( 1 - \sqrt{\frac{a_{\rm in}}{a}} \right) \exp \left[- \left( \frac{a}{a_{\rm cut}} \right)^{1/2} \right],
\end{equation}
where $p=1$ and $a_{\rm in}=3\,a_{\rm bin}$, consistent with what assumed for the simulations but with the addition of an inner smoothing term, to have $\Sigma(a_{\rm in})=0$, and an exponential cutoff from $a_{\rm cut}=18\, a_{\rm bin}$, while the disc outer radius is $a_{\rm out}=24 a_{\rm bin}$.
We use a locally isothermal temperature profile, defined prescribing $c_{\rm s}=c_{\rm s0}(a/a_{\rm bin})^{-s}$, with $s=1/2$ and $c_{\rm s0}$ tuned to obtain the desired $H/R$. We model the precession effects due to the binary quadrupole potential assuming it induces a forced precession
\begin{equation}
    \dot \varpi_{\rm Q,0}=\frac{3}{4}\frac{ q}{(q+1)^{2}}\left(\frac{a}{a_{\rm bin}}\right)^{-7/2}\frac{2{\rm \pi}}{t_{\rm bin}}.
\end{equation}
We impose free boundary conditions \citep{zanazzi2020} at $a_{\rm in}=3\,a_{\rm bin}$ and $a_{\rm out}=24\,a_{\rm bin}$.
The theoretical eigenmodes are calculated assuming zero viscosity. 
We initialise our eigenmode simulations in the fundamental mode (no nodes in the eccentricity profile). 

We note as a caveat that the resulting eigenmode is intended to serve as a reasonable initial guess for a stable eccentricity profile in the numerical simulations. We do not expect the eigenmode profiles obtained using the approach described above to be fully stable at the start of the simulations. The main aspects contributing to this are: i) the simulations employ a radial temperature profile, in contrast to the semi-major axis profile used to derive the theoretical eigenmodes; ii) the vertical velocities are not initialized to match the theoretical velocities expected from the onset of the breathing mode associated with the eccentricity profile (e.g. \citealp{lynch2021,ragusa2024}); iii) the eigenmode profiles discussed in this Section and the simulations assume an initial cavity of $a_{\rm cav}^{\rm in}=a_{\rm in}=3a_{\rm bin}$ in size, which does not correspond to the truncation radius that the binary would produce; iv) Finally, the simulations include viscosity, whereas the theoretical eigenmode profiles were derived assuming inviscid conditions.

Despite these caveats, we generally observe that the simulations relax to the appropriate eccentricity profile within a few tens of orbits, confirming that the eigenmode provides a reasonable initial guess. This can be seen in the fourth row of Fig. \ref{fig:simevo}, which shows that the system precesses rigidly from the beginning of the numerical simulation and eventually settles into a stable configuration with the appropriate truncation radius.

\section{Dependence on different parameters in the $e_{\rm cav}$ vs $H/R$, $\alpha$, $e_{\rm bin}$, $q_{\rm bin}$ }\label{appendix:simpar}

\begin{figure*}[b]
   \centering   
    \includegraphics[width=0.49\textwidth]{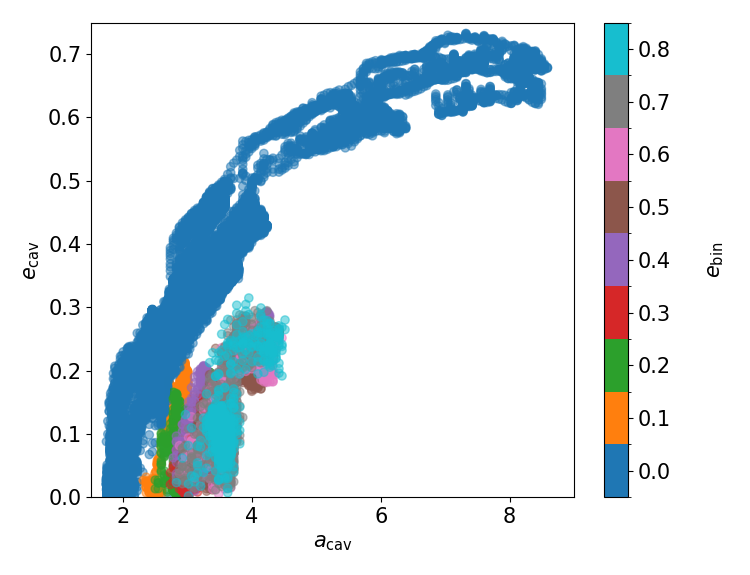}
   \includegraphics[width=0.49\textwidth]{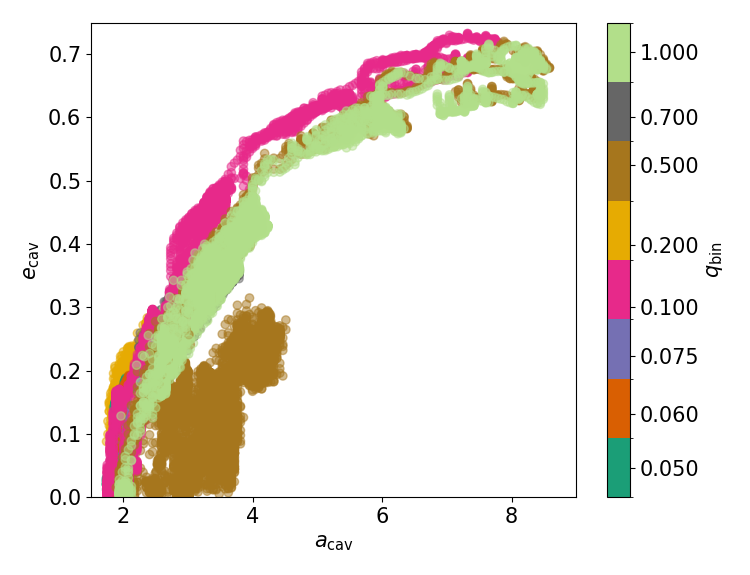}\\
   \includegraphics[width=0.49\textwidth]{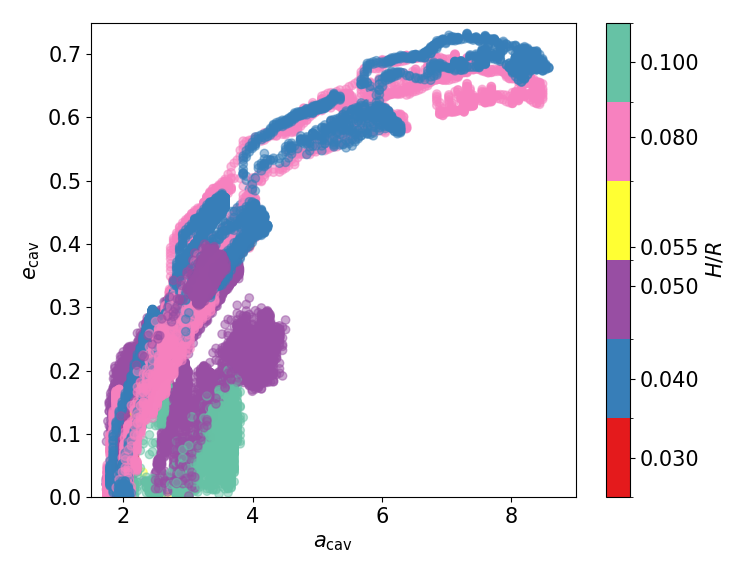}
   \includegraphics[width=0.49\textwidth]{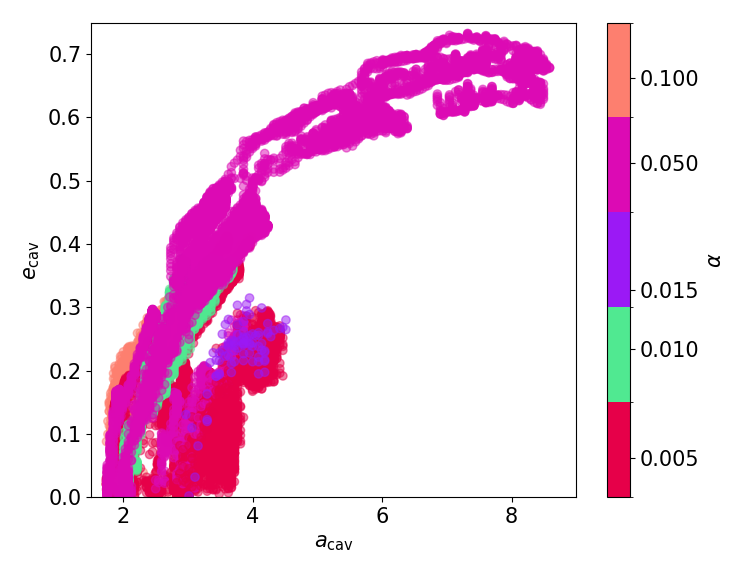}
 
   \caption{Cavity eccentricity $e_{\rm cav}$ vs cavity semi-major axis $a_{\rm cav}$ from the simulations. Same as left panel of Fig. \ref{fig:dataset} but highlighting the dependence on system parameters in different colours.}
\label{fig:dataset2}%
\end{figure*}

\begin{figure*}[b]
   \centering   
    \includegraphics[width=0.49\textwidth]{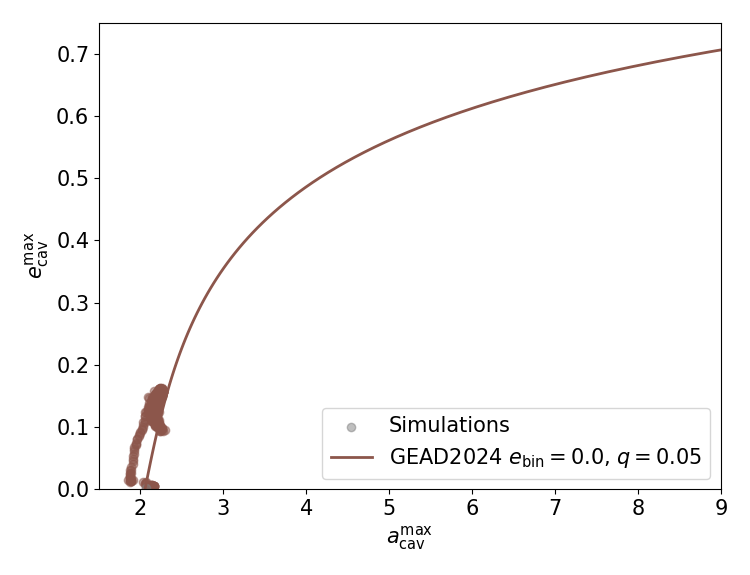}
   \includegraphics[width=0.49\textwidth]{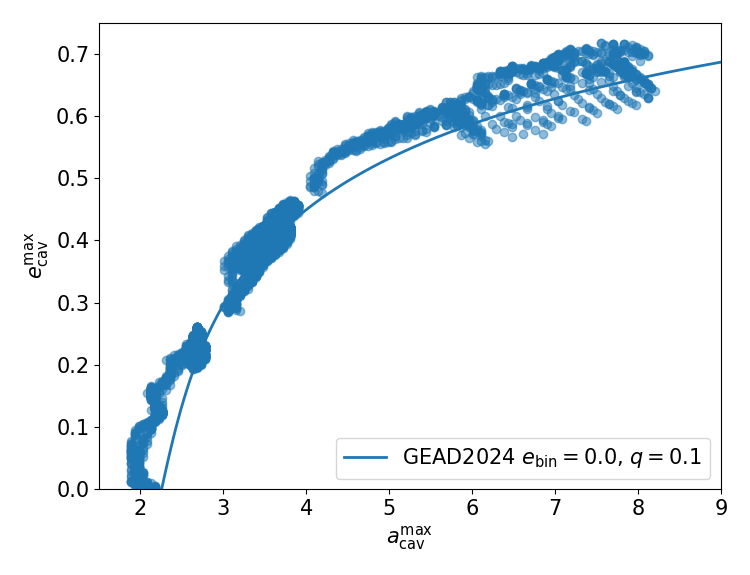}\\
   \includegraphics[width=0.49\textwidth]{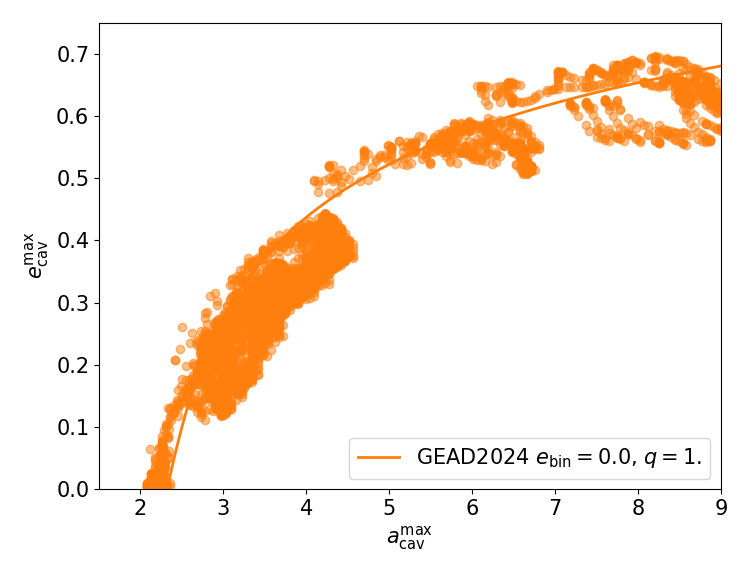}
   \includegraphics[width=0.49\textwidth]{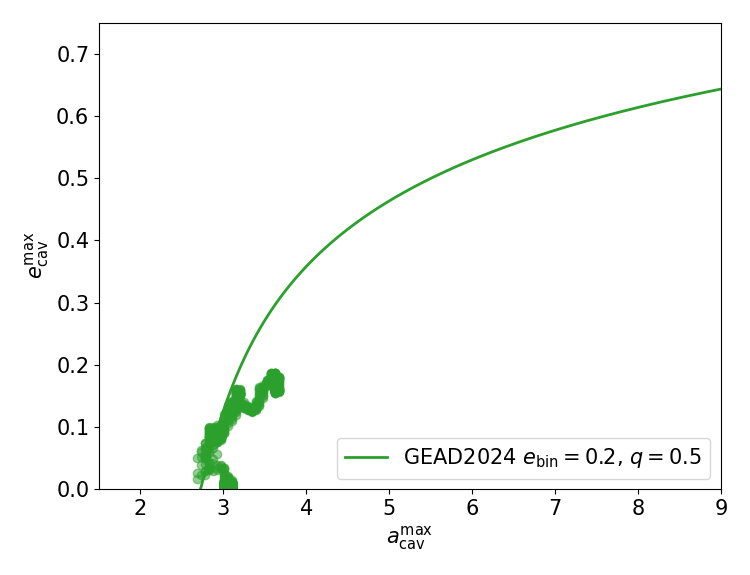}\\
   \includegraphics[width=0.49\textwidth]{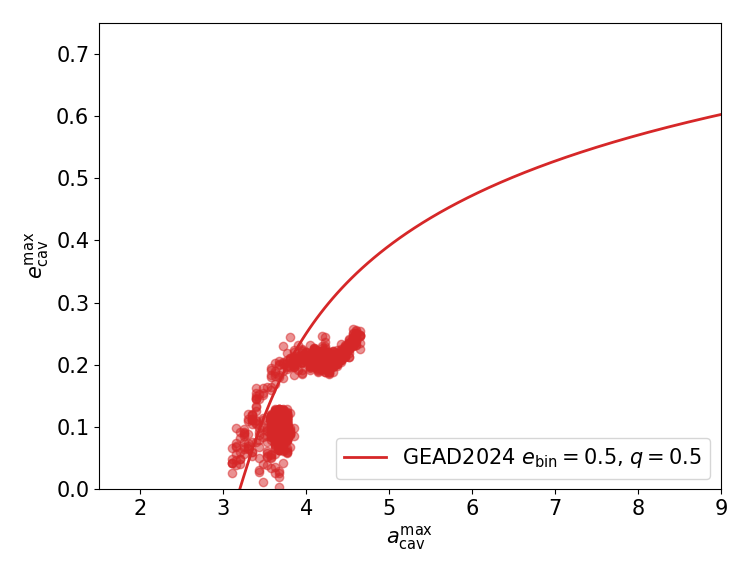}
   \includegraphics[width=0.49\textwidth]{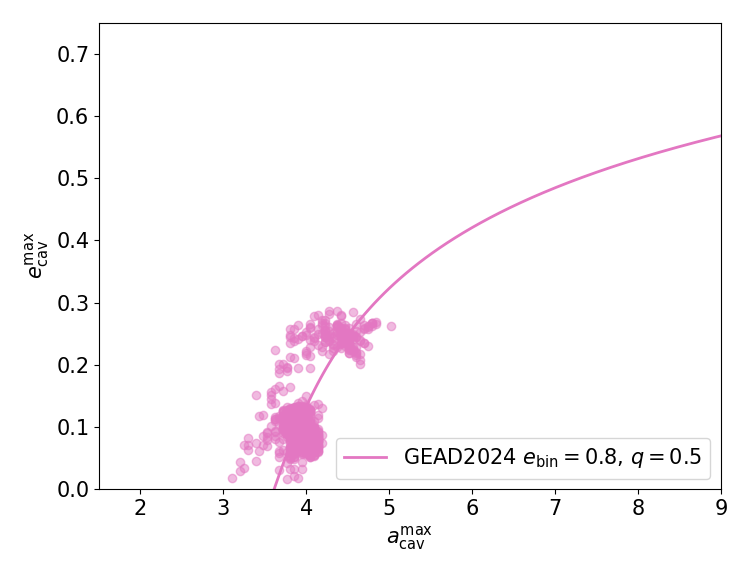}
 
   \caption{Cavity eccentricity $e_{\rm cav}^{\rm max}$ vs cavity density maximum $a_{\rm max}$ (Eq. \ref{eq:acavmax}). Each panel shows  the snapshots from simulations (data points) using $e_{\rm bin}$ and $q$ as indicated in the legend of each panel; solid curves represent cavity size predictions using the prescription for the semi-major axis of the last stable orbit around the binary from \citet{georgakarakos2024}. }
\label{fig:datasetNbody}%
\end{figure*}

We report in Tab. \ref{tab:group1}, \ref{tab:group2}, \ref{tab:group3}, \ref{tab:group4} the parameters used in the simulations.
In Fig. \ref{fig:dataset2} we mark in different colours the data points shown in Fig. \ref{fig:dataset} to highlight how different regions of the plot are associated to different parameters of the system. In Fig. \ref{fig:datasetNbody} we show a detailed comparison between the predictions for $R_{\rm p}^{\rm crit}$ from \citet{georgakarakos2024} and location of density maximum as discussed in Sec. \ref{sec:compNbody}.

Results from \ref{fig:acavecavq1}, \ref{fig:dataset2}, and \ref{fig:datasetNbody} exhibit a visible correlation with the analytical prediction from \citet{georgakarakos2024}, tracking the overall trend where more eccentric discs create cavities with larger semi-major axes. However, a closer inspection reveals that the data points do not strictly adhere to the theoretical boundary; rather, they exhibit moderate scatter both above and below the theoretical curve. In principle, given its definition of ``orbital stability limit'', it would be reasonable to expect that this curve acts as a hard wall inside which no material should be observed on closed orbits; despite this, such deviations appears to be common.

We believe it is important to stress two considerations concerning the discrepancies described above. First, unlike the idealized test-particles modeled in pure $N$-body frameworks, the gas in our hydrodynamical simulations is subject to internal pressure gradients and viscous diffusion (see Sec. \ref{sec:hydrvisc}), allowing gas elements to populate regions that would otherwise be dynamically unstable over long orbital timescales ($\sim 10^6$ binary orbits). This effectively blurs the sharp cutoff predicted for pressureless, inviscid particles. Second, the analytical boundaries derived by \citet{georgakarakos2024} represent smooth parametric fits to an inherently chaotic phase space. In the restricted three-body problem, the transition between stable and unstable orbits mediated by mean-motion resonance overlap is highly irregular, typically characterised by a complex, 
``ragged'' or sawtooth profile (e.g., see Fig. 3 of \citealt{shevchenko2015} or Fig. 6 in \citealt{adelbert2023}). This produces a  scatter in $a_{\rm cav}$ values even for adjacent $e_{\rm cav}$ values.

\begin{table}[h!]
\caption{Group 1 simulations (previously presented in \citealp{ragusa2020}, IDs do not match). }\label{tab:group1}
\centering
\begin{tabular}{l|ccccccc}
\hline\hline
ID & $q$ & $e_{\rm bin}$ & $H/R$ & $\alpha$ & $e_{\rm d}$ & $a_{\rm cav}^{\rm in}$ & $t_{\rm fin}$ \\ 
\hline \\[-8pt]
1A$e_{0}$ & 0.2 & 0.0 & 0.05 & 0.005 & 0 & 2 & 3506 \\
1E$e_{0}$ & 0.2 & 0.0 & 0.05 & 0.1   & 0 & 2 & 4972 \\
1G$e_{0}$ & 0.2 & 0.0 & 0.05 & 0.01  & 0 & 2 & 2201 \\
1H$e_{0}$ & 0.2 & 0.0 & 0.05 & 0.01  & 0 & 2 & 2761 \\
1N$e_{0}$ & 0.2 & 0.0 & 0.03 & 0.005 & 0 & 2 & 3001 \\
1Z$e_{0}$ & 0.2 & 0.0 & 0.05 & 0.01  & 0 & 2 & 2001 \\
2A$e_{0}$ & 0.1 & 0.0 & 0.05 & 0.005 & 0 & 2 & 3001 \\
4A$e_{0}$ & 0.5 & 0.0 & 0.05 & 0.005 & 0 & 2 & 2606 \\
4E30$e_{0}$ & 0.5 & 0.0 & 0.055 & 0.1 & 0 & 2 & 785 \\
4FF$e_{0}$ & 0.5 & 0.4 & 0.05 & 0.005 & 0 & 2 & 2431 \\
4F$e_{0}$ & 0.5 & 0.1 & 0.05 & 0.005 & 0 & 2 & 2661 \\
4N$e_{0}$ & 0.5 & 0.0 & 0.03 & 0.005 & 0 & 2 & 840 \\
4Z$e_{0}$ & 0.5 & 0.0 & 0.05 & 0.01 & 0 & 2 & 2481 \\
5A$e_{0}$ & 0.7 & 0.0 & 0.05 & 0.005 & 0 & 2 & 3001 \\
5E$e_{0}$ & 0.7 & 0.0 & 0.05 & 0.1   & 0 & 2 & 3001 \\
5Z$e_{0}$ & 0.7 & 0.0 & 0.05 & 0.01  & 0 & 2 & 2231 \\
6A$e_{0}$ & 1.0 & 0.0 & 0.05 & 0.005 & 0 & 2 & 5939 \\
6E$e_{0}$ & 1.0 & 0.0 & 0.05 & 0.1   & 0 & 2 & 2716 \\
6Z$e_{0}$ & 1.0 & 0.0 & 0.05 & 0.01  & 0 & 2 & 3001 \\
7A$e_{0}$ & 0.05 & 0.0 & 0.05 & 0.005 & 0 & 2 & 5106 \\
8A$e_{0}$ & 0.075 & 0.0 & 0.05 & 0.005 & 0 & 2 & 3001 \\
9A$e_{0}$ & 0.06 & 0.0 & 0.05 & 0.005 & 0 & 2 & 3001 \\ 
\hline\hline
\end{tabular}
\tablefoot{Notes: $q$ is the binary mass ratio; $e_{\rm bin}$ is the binary eccentricity; $H/R$ is the disc aspect ratio (constant across the entire disc); $\alpha$ is the \citet{shakura1973} viscous parameter, $e_{\rm d}$ is the initial disc eccentricity; $a_{\rm cav}^{\rm in}$ is the initial cavity size; $t_{\rm fin}$ is the simulation ending time in binary orbital times $t_{\rm bin}$. Simulations from this group have their IDs ending with $e_{0}$.}
\end{table}

\begin{table}[h!]
\caption{Group 2 simulations.}\label{tab:group2}
\centering
\begin{tabular}{l|ccccccc}
\hline\hline
ID & $q$ & $e_{\rm bin}$ & $H/R$ & $\alpha$ & $e_{\rm d}$ & $a_{\rm cav}^{\rm in}$ & $t_{\rm fin}$ \\ 
\hline \\[-8pt]
1A$e_{\rm b}$ & 0.5 & 0.0 & 0.1  & 0.005 & 0 & 2 & 1450 \\
2A$e_{\rm b}$ & 0.5 & 0.1 & 0.1  & 0.005 & 0 & 2 & 1630 \\
3A$e_{\rm b}$ & 0.5 & 0.2 & 0.1  & 0.005 & 0 & 2 & 1715 \\
3B$e_{\rm b}$ & 0.5 & 0.2 & 0.05 & 0.005 & 0 & 2 & 592 \\
3D$e_{\rm b}$ & 0.5 & 0.2 & 0.05 & 0.05  & 0 & 2 & 226 \\
4A$e_{\rm b}$ & 0.5 & 0.3 & 0.1  & 0.005 & 0 & 2 & 1730 \\
4B$e_{\rm b}$ & 0.5 & 0.3 & 0.05 & 0.005 & 0 & 2 & 448 \\
5A$e_{\rm b}$ & 0.5 & 0.4 & 0.1  & 0.005 & 0 & 2 & 1691 \\
5D$e_{\rm b}$ & 0.5 & 0.4 & 0.05 & 0.05  & 0 & 2 & 143 \\
6A$e_{\rm b}$ & 0.5 & 0.5 & 0.1  & 0.005 & 0 & 2 & 1513 \\
6B$e_{\rm b}$ & 0.5 & 0.5 & 0.05 & 0.005 & 0 & 2 & 507 \\
7A$e_{\rm b}$ & 0.5 & 0.6 & 0.1  & 0.005 & 0 & 2 & 1441 \\
7B$e_{\rm b}$ & 0.5 & 0.6 & 0.05 & 0.005 & 0 & 2 & 376 \\
7D$e_{\rm b}$ & 0.5 & 0.7 & 0.05 & 0.05  & 0 & 2 & 143 \\
8A$e_{\rm b}$ & 0.5 & 0.7 & 0.1  & 0.005 & 0 & 2 & 1610 \\
8B$e_{\rm b}$ & 0.5 & 0.7 & 0.05 & 0.005 & 0 & 2 & 201 \\
9A$e_{\rm b}$ & 0.5 & 0.8 & 0.1  & 0.005 & 0 & 2 & 1923 \\
9B$e_{\rm b}$ & 0.5 & 0.8 & 0.05 & 0.005 & 0 & 2 & 333 \\
9D$e_{\rm b}$ & 0.5 & 0.8 & 0.05 & 0.015 & 0 & 2 & 273 \\
\hline\hline
\end{tabular}
\tablefoot{Notes: See the Notes of Table \ref{tab:group1} for a description of the columns. Simulations from this group have their ID ending with $e_{\rm b}$.}
\end{table}

\begin{table}[h!]
\caption{Group 3 simulations. }\label{tab:group3}
\centering
\begin{tabular}{l|ccccccc}
\hline\hline 
ID & $q$ & $e_{\rm bin}$ & $H/R$ & $\alpha$ & $e_{\rm d}$ & $a_{\rm cav}^{\rm in}$ & $t_{\rm fin}$ \\
\hline\\[-8pt]
01A0$e_{\rm d}$ & 0.1 & 0.0 & 0.08 & 0.05 & 0.0  & 2 & 722 \\
01A3$e_{\rm d}$ & 0.1 & 0.0 & 0.08 & 0.05 & 0.3  & 3 & 497 \\
01A5$e_{\rm d}$ & 0.1 & 0.0 & 0.08 & 0.05 & 0.5  & 4 & 253 \\
01A6$e_{\rm d}$ & 0.1 & 0.0 & 0.08 & 0.05 & 0.65 & 6 & 357 \\
01B0$e_{\rm d}$ & 0.1 & 0.0 & 0.04 & 0.05 & 0.0  & 2 & 1000 \\
01B3$e_{\rm d}$ & 0.1 & 0.0 & 0.04 & 0.05 & 0.3  & 3 & 822 \\
01B5$e_{\rm d}$ & 0.1 & 0.0 & 0.04 & 0.05 & 0.5  & 4 & 231 \\
01B6$e_{\rm d}$ & 0.1 & 0.0 & 0.04 & 0.05 & 0.65 & 6 & 218 \\
05A0$e_{\rm d}$ & 0.5 & 0.0 & 0.08 & 0.05 & 0.0  & 2 & 571 \\
05A3$e_{\rm d}$ & 0.5 & 0.0 & 0.08 & 0.05 & 0.3  & 3 & 517 \\
05A5$e_{\rm d}$ & 0.5 & 0.0 & 0.08 & 0.05 & 0.5  & 4 & 161 \\
05A6$e_{\rm d}$ & 0.5 & 0.0 & 0.08 & 0.05 & 0.65 & 6 & 243 \\
05B0$e_{\rm d}$ & 0.5 & 0.0 & 0.04 & 0.05 & 0.0  & 2 & 558 \\
05B3$e_{\rm d}$ & 0.5 & 0.0 & 0.04 & 0.05 & 0.3  & 3 & 507 \\
05B5$e_{\rm d}$ & 0.5 & 0.0 & 0.04 & 0.05 & 0.5  & 4 & 300 \\
05B6$e_{\rm d}$ & 0.5 & 0.0 & 0.04 & 0.05 & 0.65 & 6 & 220 \\
1A0$e_{\rm d}$ & 1.0 & 0.0 & 0.08 & 0.05 & 0.0  & 2 & 514 \\
1A3$e_{\rm d}$ & 1.0 & 0.0 & 0.08 & 0.05 & 0.3  & 3 & 484 \\
1A5$e_{\rm d}$ & 1.0 & 0.0 & 0.08 & 0.05 & 0.5  & 4 & 157 \\
1A6$e_{\rm d}$ & 1.0 & 0.0 & 0.08 & 0.05 & 0.65 & 6 & 299 \\
1B0$e_{\rm d}$ & 1.0 & 0.0 & 0.04 & 0.05 & 0.0  & 2 & 631 \\
1B2$e_{\rm d}$ & 1.0 & 0.0 & 0.05 & 0.05 & 0.2  & 2 & 931 \\ 
1B3$e_{\rm d}$ & 1.0 & 0.0 & 0.04 & 0.05 & 0.3  & 3 & 632 \\
1B5$e_{\rm d}$ & 1.0 & 0.0 & 0.04 & 0.05 & 0.5  & 4 & 341 \\
1B6$e_{\rm d}$ & 1.0 & 0.0 & 0.04 & 0.05 & 0.65 & 6 & 321 \\
\hline\hline
1A0$e_{\rm d,0}$ & 1.0 & 0.0 & 0.1 & 0.05 & 0.0  & 2 & 139 \\
1A2$e_{\rm d,0}$ & 1.0 & 0.0 & 0.1 & 0.05 & 0.2  & 2 & 196 \\
1A4$e_{\rm d,0}$ & 1.0 & 0.0 & 0.1 & 0.05 & 0.4  & 2 & 124 \\
1A5$e_{\rm d,0}$ & 1.0 & 0.0 & 0.1 & 0.05 & 0.5 & 2 & 132 \\
1B0$e_{\rm d,0}$ & 1.0 & 0.0 & 0.05 & 0.05 & 0.0  & 2 & 143 \\
1B2$e_{\rm d,0}$ & 1.0 & 0.0 & 0.05 & 0.05 & 0.2  & 2 & 931 \\
1B4$e_{\rm d,0}$ & 1.0 & 0.0 & 0.05 & 0.05 & 0.4  & 2 & 139 \\
1B5$e_{\rm d,0}$ & 1.0 & 0.0 & 0.05 & 0.05 & 0.5 & 2 & 110 \\
\hline\hline
\end{tabular}
\tablefoot{Notes: see Notes of Tab. \ref{tab:group1} for a description of the columns. Simulations from this group have their ID ending with $e_{\rm d}$, referring to simulations with constant eccentricity profiles.}
\end{table}

\begin{table}[h!]
\caption{Group 4 simulations. }\label{tab:group4}
\centering
\begin{tabular}{l|ccccccc}
\hline\hline 
ID & $q$ & $e_{\rm bin}$ & $H/R$ & $\alpha$ & $e_{\rm d}$ & $a_{\rm cav}^{\rm in}$ & $t_{\rm fin}$ \\
\hline\\[-8pt]
01A3$e_{\rm d}^{\rm m}$ & 0.1 & 0.0 & 0.08 & 0.01 & 0.3 & 3 & 3128 \\
01B3$e_{\rm d}^{\rm m}$ & 0.1 & 0.0 & 0.04 & 0.01 & 0.3 & 3 & 1588 \\
05A3$e_{\rm d}^{\rm m}$ & 0.5 & 0.0 & 0.08 & 0.01 & 0.3 & 3 & 3401 \\
05B3$e_{\rm d}^{\rm m}$ & 0.5 & 0.0 & 0.04 & 0.01 & 0.3 & 3 & 1639 \\
1A3$e_{\rm d}^{\rm m}$ & 1.0 & 0.0 & 0.08 & 0.01 & 0.3 & 3 & 1497 \\
1B3$e_{\rm d}^{\rm m}$ & 1.0 & 0.0 & 0.04 & 0.01 & 0.3 & 3 & 2041 \\ 
\hline\hline
\end{tabular}
\tablefoot{Notes: see the Notes of Tab. \ref{tab:group1} for a description of the columns. Simulations from this group have their ID ending with $e_{\rm d}^{\rm m}$, referring to simulations with eccentricity profiles decreasing with radius of the fundamental eccentric eigen-mode.}
\end{table}

\section{MCMC Cornerplot}\label{appendix:MCMC}

We obtain the values of $c_1\textrm{--}c_9$ in Eq. (\ref{eq:acav}) using Bayesian inference by running an MCMC analysis on the dataset. We adopt a Gaussian likelihood with standard deviation $\sigma_0$, which is inferred as a free parameter:
\begin{equation}
    \log \mathcal L= -N\log\left(\sqrt{2{\rm \pi}}\sigma_0\right) +\sum_j^N -\frac{1}{2}\frac{\left(R_{{\rm p},j}-R_{{\rm p},j}^{\rm th}\right)^2}{\sigma_0^2}
\end{equation}
where $R_{{\rm p},j}=a_{{\rm cav},j}(1-e_{{\rm cav},j})$, is the pericentre radius of the $j$-th simulation snapshot in our dataset, $R_{{\rm p},j}^{\rm th}$ is the corresponding pericentre radius predicted using Eq. (\ref{eq:acav}), and $N$ is the total number of simulation snapshots. We assume uniform priors $-5\leq c\leq 5$ for all coefficients, and run the MCMC for $2\times 10^4$ steps with 20 walkers, discarding the first $17000$ steps.

\begin{figure*}
   \centering   \includegraphics[width=\textwidth]{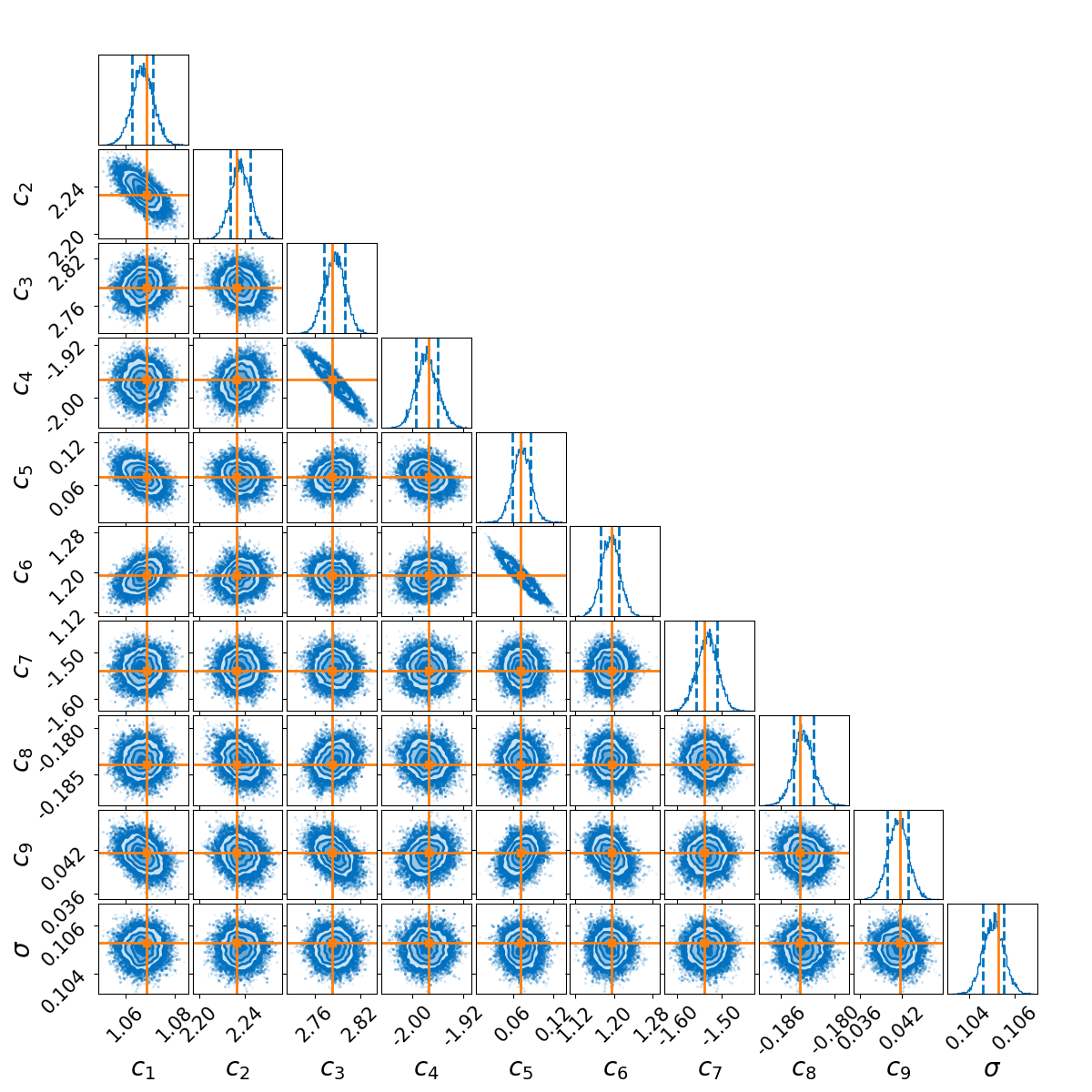}
   \caption{Posterior distributions for the parameters $c_1\textrm{--}c_9$ in Eq. (\ref{eq:acav}). A summary of the most likely values is reported in Tab. \ref{tab:MCMC}.}
\label{fig:cornerplot}%
\end{figure*}

We report in Tab. \ref{tab:MCMC} the values of the parameters $c_1\textrm{--}\,c_9$ and $\sigma_0$ corresponding to the maximum likelihood from the MCMC procedure.
Figure \ref{fig:cornerplot} shows posterior distributions for $c_1\textrm{--}c_9$ parameters in Eq. (\ref{eq:acav}).

\end{document}